\documentclass[journal]{IEEEtran}
\ifCLASSINFOpdf
\else
\fi

\usepackage{cite}
\usepackage{amsmath,amssymb,amsfonts}
\usepackage{algorithmic}
\usepackage{graphicx}
\usepackage{textcomp}
\usepackage{xcolor}
\usepackage{pgfplots, filecontents}
\usepgfplotslibrary{patchplots}
\usepackage{enumerate}
\usepackage{multirow}
\usepackage[nohyperlinks, nolist]{acronym}
\usepackage{epstopdf}
\usepackage{float}
\usepackage[caption = false]{subfig}
\usetikzlibrary{matrix}
\graphicspath{{./fig/}}
\def\BibTeX{{\rm B\kern-.05em{\sc i\kern-.025em b}\kern-.08em
    T\kern-.1667em\lower.7ex\hbox{E}\kern-.125emX}}

\usepackage{tikz, pgfplots}
\pgfplotsset{compat=1.16}
\usetikzlibrary{positioning}
\usetikzlibrary{arrows}
\usetikzlibrary{shapes.multipart}
\usetikzlibrary{calc,through,backgrounds}
\usepackage[margin=2cm]{geometry}
      
  \usepackage{array}
  \newcolumntype{R}{>{$}r<{$}}
  \newcolumntype{L}{>{$}l<{$}}
  \newcolumntype{M}{R@{${}={}$}L}
  
\usepackage{scalerel}[2016/12/29]
\newcommand\Largerightarrow{\hstretch{2}{\rightarrow}}
 \newcommand\Largeleftarrow{\hstretch{2}{\leftarrow}}     
      
\newcommand{\mynearrow}{\mathrel{\rotatebox[origin=c]{45}{$\rightarrow$}}}

\newcommand{\mysearrow}{\mathrel{\rotatebox[origin=c]{-45}{$\rightarrow$}}}
   
\newcommand{\myswarrow}{\mathrel{\rotatebox[origin=c]{-135}{$\rightarrow$}}} 

\newcommand{\mynwarrow}{\mathrel{\rotatebox[origin=c]{135}{$\rightarrow$}}} 
    
\newcommand{\ve}{\mathbf}
\newcommand{\m}{\mathbf}

\newcommand{\vef}[1]{\mathbf{\tilde{\mathbf{#1}}}} 

\newcommand{\veh}[1]{\widehat{\mathbf{#1}}}
\newcommand{\mh}[1]{\widehat{\mathbf{#1}}}

\newcommand*{\ReadOutElement}[4]{%
    \pgfplotstablegetelem{#2}{#3}\of{#1}%
    \let#4\pgfplotsretval
}

\begin{document}
%


\begin{acronym}

\acro{accdm}[AC-CDM]{auto-correlation-based code-division multiplexing}
\acro{ad}[AD]{autonomous driving}
\acro{adc}[ADC]{analog-to-digital converter}
\acro{adas}[ADAS]{advanced driver assistance systems}
\acro{als}[ALS]{approximate least squares}
\acro{aoa}[AOA]{angle of arrival}
\acro{aod}[AOD]{angle of departure}
\acro{awgn}[AWGN]{additive white Gaussian noise}
\acro{bb}[BB]{baseband}
\acro{ber}[BER]{bit error ratio}
\acro{blue}[BLUE]{best linear unbiased estimator}
\acro{bmse}[BMSE]{Bayesian mean square error}
\acro{bpsk}[BPSK]{binary phase shift keying}
\acro{bwlue}[BWLUE]{best widely linear unbiased estimator}
\acro{cfo}[CFO]{carrier frequency offset}
\acro{cfr}[CFR]{channel frequency response}
\acro{cir}[CIR]{channel impulse response}
\acro{cpe}[CPE]{common phase error}
\acro{cs}[CS]{compressed sensing}
\acro{cwcu}[CWCU]{component-wise conditionally unbiased}
\acro{cl}[CWCU LMMSE]{component-wise conditionally unbiased linear minimum mean square error}
\acro{cp}[CP]{cyclic prefix}
\acro{cwl}[CWCU WLMMSE]{component-wise conditionally unbiased widely linear minimum mean square error}
\acro{dbf}[DBF]{digital beamforming}
\acro{dc}[DC]{direct current}
\acro{dft}[DFT]{discrete Fourier transform}
\acro{ddm}[DDM]{Doppler-division multiplexing}
\acro{dpsk}[DPSK]{differential phase shift keying}
\acro{dsi}[NeqDySI]{non-equidistant dynamic subcarrier interleaving}
\acro{ecir}[ECIR]{effective channel impulse response}
\acro{ecfr}[ECFR]{effective channel frequency response}
\acro{em}[EM]{expectation-maximization}
\acro{esi}[ESI]{equidistant subcarrier interleaving}
\acro{fmcw}[FMCW]{frequency-modulated continuous wave}
\acro{fft}[FFT]{fast Fourier transform}
\acro{fir}[FIR]{finite impulse response}
\acro{ici}[ICI]{inter-carrier interference}
\acro{isi}[ISI]{inter-symbol interference}
\acro{idft}[IDFT]{inverse discrete Fourier transform}
\acro{ifft}[IFFT]{inverse fast Fourier transform}
\acro{iid}[i.i.d.]{independent and identically distributed}
\acro{llr}[LLR]{log-likelihood ratio}
\acro{lmmse}[LMMSE]{linear minimum mean square error}
\acro{lms}[LMS]{least mean square}
\acro{los}[LOS]{line of sight}
\acro{ls}[LS]{least squares}
\acro{lti}[LTI]{linear time-invariant}
\acro{map}[MAP]{maximum a posteriori}
\acro{mimo}[MIMO]{multiple-input multiple-output}
\acro{miso}[MISO]{multiple-input single-output}
\acro{ml}[ML]{maximum likelihood}
\acro{mlem}[ML-EM]{maximum likelihood expectation-maximization}
\acro{mmse}[MMSE]{minimum mean square error}
\acro{mrscdm}[MRS-CDM]{modified repeated symbol CDM}
\acro{mse}[MSE]{mean square error}
\acro{mvdr}[MVDR]{minimum variance distortionless response}
\acro{mvu}[MVU]{minimum variance unbiased}
\acro{nlms}[NLMS]{normalized least mean squares}
\acro{nlos}[NLOS]{non-line of sight}
\acro{ofdm}[OFDM]{orthogonal frequency-division multiplexing}
\acro{otfs}[OTFS]{orthogonal time frequency space}
\acro{pdf}[PDF]{probability density function}
\acro{pwcu}[PWCU]{part-wise conditionally unbiased}
\acro{pwl}[PWCU WLMMSE]{part-wise conditionally unbiased widely linear minimum mean square error}
\acro{qam}[QAM]{quadrature amplitude modulation}
\acro{qpsk}[QPSK]{quadrature phase-shift keying}
\acro{rcs}[RCS]{radar cross section}
\acro{rls}[RLS]{recursive least squares}
\acro{rvm}[RDM]{range-Doppler map}
\acro{rdm}[RDMult]{range-division multiplexing}
\acro{sim}[SIM]{spectral interleaving multiplexing}
\acro{siso}[SISO]{single-input single-output}
\acro{snr}[SNR]{signal-to-noise ratio}
\acro{stln}[STLN]{structured total least norm}
\acro{stls}[STLS]{structured total least squares}
\acro{tls}[TLS]{total least squares}
\acro{tdm}[TDM]{time-division multiplexing}
\acro{ula}[ULA]{uniform linear array}
\acro{uwofdm}[UW-OFDM]{unique-word orthogonal frequency division multiplexing}
\acro{wlan}[WLAN]{wireless local area network}
\acro{wlls}[WLLS]{widely linear least squares}
\acro{wlmmse}[WLMMSE]{widely linear minimum mean square error}
\acro{wls}[WLS]{weighted least squares}
\acro{wrt}[w.r.t.]{with respect to}
\acro{wwlls}[WWLLS]{weighted widely linear least squares}

\end{acronym}

\title{Doppler-Division Multiplexing for MIMO OFDM Joint Sensing and Communications}
%
%
%

\author{Oliver~Lang,~\IEEEmembership{Member,~IEEE,}
		Christian~Hofbauer,~\IEEEmembership{Member,~IEEE,}
        Reinhard~Feger,         
        and~Mario~Huemer,~\IEEEmembership{Senior~Member,~IEEE}
\thanks{Oliver Lang is with the Institute
of Signal Processing, Johannes Kepler University, Linz,
Austria (e-mail: oliver.lang@jku.at).}
\thanks{Christian Hofbauer is with Silicon Austria Labs GmbH, Linz, Austria (e-mail: christian.hofbauer@silicon-austria.com).}
\thanks{Reinhard Feger is with the Institute for Communications Engineering and RF-Systems, Johannes Kepler University, Linz,
Austria (e-mail: reinhard.feger@jku.at).}
\thanks{Mario Huemer is with the Institute
of Signal Processing, Johannes Kepler University, Linz,
Austria, and also with the JKU LIT SAL eSPML Lab, 4040 Linz, Austria (e-mail: mario.huemer@jku.at).}
\thanks{ The work of Christian Hofbauer was supported by Silicon Austria Labs (SAL), owned by the Republic of Austria, the Styrian Business Promotion Agency (SFG), the federal state of Carinthia, the Upper Austrian Research (UAR), and the Austrian Association for the Electric and Electronics Industry (FEEI). 

This work was supported by the "University SAL Labs" initiative of SAL and its Austrian partner universities for applied fundamental research for electronic based systems. }
}

\maketitle

\begin{abstract}
A promising waveform candidate for future joint sensing and communication systems is \ac{ofdm}. For such systems, supporting multiple transmit antennas requires multiplexing methods for the generation of orthogonal transmit signals, where \ac{esi} is the most popular multiplexing method. 
 In this work, we analyze a multiplexing method called \ac{ddm}. This method applies a phase shift from \ac{ofdm} symbol to \ac{ofdm} symbol to separate signals transmitted by different Tx antennas along the velocity axis of the range-Doppler map. While general properties of \ac{ddm} for the task of radar sensing are analyzed in this work, the main focus lies on the implications of \ac{ddm} on the communication task. It will be shown that for \ac{ddm}, the channels observed in the communication receiver are heavily time-varying, preventing any meaningful transmission of data when not taken into account. In this work, a communication system  designed to combat these time-varying channels is proposed, which includes methods for data estimation, synchronization, and channel estimation.
\Ac{ber} simulations demonstrate the superiority of this communications system compared to a system utilizing \ac{esi}. 
\end{abstract}


\acresetall

%
\IEEEpeerreviewmaketitle

\section{Introduction}  \label{sec:Introduction}

\IEEEPARstart{P}{otential}  applications of systems capable of joint sensing and communications include automotive car-to-car communications and cellular sensing \cite{akan2020internet, ma2020joint, ADAS2,Gerstmair1ADAS, ADAS1, Gerstmair2ADAS}. While many different system architectures and waveform designs are possible, a so-called dual-function radar-communication system is assumed in this work that uses the very same transmit signals for both, radar sensing and  communications, simultaneously. Further, it is assumed that the communication receiver and the radar receiver are located at different positions.

A prominent waveform for joint sensing and communication systems is \ac{ofdm}  \cite{Levanon_Multifrequency_complementary, Donnet_Combining_MIMO_Radar, Sturm_A_novel_approach, Garmatyuk_Feasibility_study, Sturm_An_OFDM_System, Sturm_Performance_verification, Sit_Automotive_MIMO_OFDM,Braun_Parametrization, Hakobyan_A_novel_OFDM_MIMO, Hakobyan_Inter_Carrier_Interference, lang2022ofdm}, which is also basis for the investigations in this work.

For many radar sensing applications, detecting the angular positions of objects in the sensor's vicinity is of importance. This is usually achieved by utilizing \ac{dbf} in combination with several transmit (Tx) and receive (Rx) antennas. These so-called \ac{mimo} systems \cite{8443533} employ multiplexing methods for generating orthogonal transmit signals that are separable in the receiver. 
The most popular multiplexing method for the \ac{ofdm} waveform is \ac{esi}  \cite{Sit_Automotive_MIMO_OFDM, sturm2013spectrally}, for which each subcarrier is assigned to only one of the $N_\text{Tx}$ Tx antennas (cf. Fig.~\ref{fig_MIMO_spectral_interleaving} a). Hence, signals radiated by different Tx antennas can be separated in frequency domain. Several extensions of \ac{esi} exist with randomly allocated subcarriers \cite{Knill_Random_Multiplexing}, with non-equidistantly allocated subcarriers \cite{Hakobyan_A_novel_OFDM_MIMO},  and with dynamically allocated subcarriers \cite{Hakobyan_A_novel_OFDM_MIMO_wo_CS}. The latter one is referred to as \ac{dsi}, and it changes the allocation of the subcarriers onto the Tx antennas from \ac{ofdm} symbol to \ac{ofdm} symbol.


\input{./fig/Fig_1_MIMO_spectral_interleaving_DDM}

A multiplexing method analyzed in \cite{knill2021coded} is denoted as \ac{accdm}. For this method, the same data are radiated on every Tx antenna and on every subcarrier except for antenna-specific time delays. These time delays, which are implemented via linearly increasing phase rotations along the subcarriers, move the corresponding receive signals along the range axis of the \ac{rvm} such that $N_\text{Tx}$ peaks appear along the range axis for each real object.

Another multiplexing method analyzed in \cite{knill2021coded} is denoted as \ac{mrscdm}, which applies orthogonal Hadamard codes onto the \ac{ofdm} symbols transmitted by different Tx antennas. In the receiver, $N_\text{Tx}$ \acp{rvm} are evaluated separately, one for each Tx antenna. In every \ac{rvm} there appears one main peak and $N_\text{Tx}-1$ spurs along the velocity axis for each real object such that the maximum unambiguous velocity is reduced by a factor of $N_\text{Tx}$.

A \ac{dft}-coded multiplexing method investigated in \cite{suh2021time} applies a \ac{dft} matrix as precoding matrix onto the transmit signals that shifts the corresponding receive signals either along the range axis, the velocity axis, or both of them. 

The multiplexing method analyzed in \cite{Lang_RDM_JP} is referred to as \ac{rdm}, and it applies a phase shift from subcarrier to subcarrier to shift the signal components of the \ac{rvm} along the range axis. The transmit signals generated by \ac{rdm} coincide with that generated by \ac{accdm} \cite{knill2021coded} and the \ac{dft}-coded multiplexing method \cite{suh2021time} for special parametrization \cite{Lang_RDM_JP}.

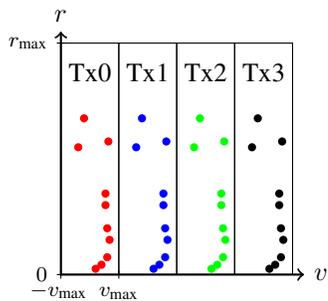
\begin{figure}[!t]
\centering
\begin{tikzpicture}[scale=0.77]
%
    %
    \begin{scope}[shift={(5.3,0)}]
    \draw [<->,thick] (0,4.2) node (yaxis) [above] {$r$}
        |- (4.2,0) node (xaxis) [right] {$v$};
    \draw[draw = black] (2pt,0) -- (-2pt,0) node[left] {\footnotesize $0$};
    \draw[draw = black] (2pt,4) -- (-2pt,4) node[left] {\footnotesize $r_\text{max}$};

    \draw[draw=black] (0,0) rectangle ++(1,4);
    \draw[draw=black] (1,0) rectangle ++(1,4);
    \draw[draw=black] (2,0) rectangle ++(1,4);
    \draw[draw=black] (3,0) rectangle ++(1,4);
    \node[align = center] at (0.5,3.5) {Tx0};
    \node[align = center] at (1.5,3.5) {Tx1};
    \node[align = center] at (2.5,3.5) {Tx2};
    \node[align = center] at (3.5,3.5) {Tx3};
	\fill[red] (1.8-1.5,2.2) circle (2pt);
    \fill[red] (1.9-1.5,2.7) circle (2pt);
    \fill[red] (2.1-1.5,0.1) circle (2pt);
    \fill[red] (2.2-1.5,0.17) circle (2pt);
    \fill[red] (2.3-1.5,0.3) circle (2pt);
    \fill[red] (2.34-1.5,0.6) circle (2pt);
    \fill[red] (2.3-1.5,0.8) circle (2pt);
    \fill[red] (2.27-1.5,1.2) circle (2pt);
    \fill[red] (2.27-1.5,1.4) circle (2pt);
    \fill[red] (2.32-1.5,2.3) circle (2pt);
    
    \fill[blue] (1.8-0.5,2.2) circle (2pt);
    \fill[blue] (1.9-0.5,2.7) circle (2pt);
    \fill[blue] (2.1-0.5,0.1) circle (2pt);
    \fill[blue] (2.2-0.5,0.17) circle (2pt);
    \fill[blue] (2.3-0.5,0.3) circle (2pt);
    \fill[blue] (2.34-0.5,0.6) circle (2pt);
    \fill[blue] (2.3-0.5,0.8) circle (2pt);
    \fill[blue] (2.27-0.5,1.2) circle (2pt);
    \fill[blue] (2.27-0.5,1.4) circle (2pt);
    \fill[blue] (2.32-0.5,2.3) circle (2pt);
    
	\fill[green] (1.8+0.5,2.2) circle (2pt);
    \fill[green] (1.9+0.5,2.7) circle (2pt);
    \fill[green] (2.1+0.5,0.1) circle (2pt);
    \fill[green] (2.2+0.5,0.17) circle (2pt);
    \fill[green] (2.3+0.5,0.3) circle (2pt);
    \fill[green] (2.34+0.5,0.6) circle (2pt);
    \fill[green] (2.3+0.5,0.8) circle (2pt);
    \fill[green] (2.27+0.5,1.2) circle (2pt);
    \fill[green] (2.27+0.5,1.4) circle (2pt);
    \fill[green] (2.32+0.5,2.3) circle (2pt);

    \fill[black] (1.8+1.5,2.2) circle (2pt);
    \fill[black] (1.9+1.5,2.7) circle (2pt);
    \fill[black] (2.1+1.5,0.1) circle (2pt);
    \fill[black] (2.2+1.5,0.17) circle (2pt);
    \fill[black] (2.3+1.5,0.3) circle (2pt);
    \fill[black] (2.34+1.5,0.6) circle (2pt);
    \fill[black] (2.3+1.5,0.8) circle (2pt);
    \fill[black] (2.27+1.5,1.2) circle (2pt);
    \fill[black] (2.27+1.5,1.4) circle (2pt);
    \fill[black] (2.32+1.5,2.3) circle (2pt);
    
    \draw[draw = black] (1,2pt) -- (1,-2pt);
    \draw[draw = black] (0,2pt) -- (0,-2pt);
    
    \node at (-1pt,-8pt) {\footnotesize $-v_\text{max}$};
	\node at (1,-9pt) {\footnotesize $v_\text{max}$};

  \end{scope}
\end{tikzpicture}
\caption{ Sketch of an RDM for DDM with $N_\text{Tx}=4$ transmit antennas. Every real object results in $N_\text{Tx}$ peaks in the RDM, each one associated with one Tx antenna. }
\label{fig_DDM_DDM}
\end{figure}

In this work, we analyze a multiplexing method referred to as \ac{ddm}. \ac{ddm} shares some similarities with the \ac{dft}-coded multiplexing method \cite{suh2021time}, with \ac{mrscdm} \cite{knill2021coded}, and with \ac{rdm} \cite{Lang_RDM_JP}, however, there exist distinct differences in some details that will be discussed later. 
\ac{ddm} modifies the transmit signal for each Tx antenna such that the received signal components are shifted along the velocity axis. A proper modification to achieve this is a Tx antenna specific phase shift from \ac{ofdm} symbol to \ac{ofdm} symbol as indicated in  Fig.~\ref{fig_MIMO_spectral_interleaving} b). This phase shift is referred to as $\Delta \psi_k$, with the Tx antenna index $k=0,1, \hdots, N_\text{Tx}-1$. 
A schematic \ac{rvm} for a \ac{mimo} \ac{ofdm} radar system utilizing \ac{ddm} is sketched in Fig.~\ref{fig_DDM_DDM} for $N_\text{Tx} = 4$ Tx antennas. This figure shows a possible alignment of the Tx antennas and their corresponding signal components within the \ac{rvm}. 
For radar sensing, the performance in terms of the \ac{snr} in the \ac{rvm} of a \ac{mimo} \ac{ofdm} radar system utilizing \ac{ddm} is
approximately equal to a \ac{mimo} \ac{ofdm} radar system employing \ac{esi}, which is analyzed in this work.



\ac{ddm} shares some similarities with the \ac{dft}-coded multiplexing method \cite{suh2021time}, which utilizes a \ac{dft} precoding matrix applied on the subcarriers and/or on the \ac{ofdm} symbols to shift the signal components radiated by different Tx antennas either along the range axis, the velocity axis, or both of them. This \ac{dft} precoding matrix, when applied onto full \ac{ofdm} symbols, corresponds to a phase shift from \ac{ofdm} symbol to \ac{ofdm} symbol as done for \ac{ddm}. However, the design rule for choosing the phase shift utilized in this work differs from the one employed in \cite{suh2021time} (cf. Sec.~\ref{sec.Choice_of_phi}). Moreover, no analysis of the implications of the phase shift on the communication task was carried out if \cite{suh2021time}.

 \ac{ddm} also shares some similarities with \ac{mrscdm} \cite{knill2021coded} by means of shifting signal components from different Tx antennas along the velocity axis. However, for \ac{mrscdm}, each Tx antenna repeatedly transmits the same \ac{ofdm} symbol during a whole frame of $N_\text{sym}$ \ac{ofdm} symbols, preventing efficient communications. In contrast to that, \ac{ddm} allows for efficient communications as will be demonstrated in this work. Moreover, the orthogonal Hadamard codes applied on the transmit \ac{ofdm} symbols for \ac{mrscdm} in general differ from the phase shift utilized in \ac{ddm} (cf. Sec.~\ref{sec.Choice_of_phi}).


The multiplexing methods \ac{rdm} and \ac{ddm} share some similarities, too. Both methods modify the transmit signals such that the corresponding received signals appear in different areas within the \ac{rvm}. However, they have the following differences (with a detailed explanation later in this work)
\begin{itemize}
\item \ac{rdm} applies a phase shift from subcarrier to subcarrier, \ac{ddm} applies a phase shift from \ac{ofdm} symbol to \ac{ofdm} symbol.
\item For \ac{rdm}, the effective communication channel shows a repetitive pattern with constructive/destructive interferences along the subcarriers, while the effective communication channel for \ac{ddm} shows a repetitive pattern with constructive/destructive interferences along the \ac{ofdm} symbols.
\item Both approaches add additional data redundancy to surpass the issue of the mentioned  interferences. This is done by transmitting the same information over several subcarriers for \ac{rdm}, and by transmitting the same information over several \ac{ofdm} symbols for \ac{ddm}. 
\item Naturally, these different approaches of adding additional redundancy require different methods for synchronization, channel estimation, and data estimation.
\end{itemize}

As already mentioned, this work considers so-called dual-function radar-communication systems that use the very same transmit signals for both, the radar sensing task and the communication task, simultaneously. For the latter task, the specific design of the transmit signal for \ac{ddm} affects the observed channel between transmitter and communication receiver. More specifically, it will turn out that the received signals at the communication receiver are affected by constructive or destructive interference, which is a typical effect for \ac{mimo} systems. However, the special transmit signals for \ac{ddm} cause this interference to be heavily time-varying. As a consequence, the channel coefficients may significantly change in magnitude and phase from one \ac{ofdm} symbol to the next one. Based on a comprehensive analysis of this observation, a communication system capable of dealing with these time-varying channels is proposed in this work, which includes adequate methods for synchronization, channel estimation, and data estimation. The communication system's performance is evaluated via extensive \ac{ber} simulations.
\\
\\
\emph{Organization}: 
\\ 
Sec.~\ref{sec:Basics_OFDM_Radar} introduces the general \ac{ofdm} waveform and the usual radar receiver signal processing. The \ac{ddm} method is described in Sec.~\ref{sec:Proposed_MIMO} and discussed in the context of radar sensing and compared to competitive multiplexing methods in Sec.~\ref{sec.Properties_DDM}. The proposed communication system for a \ac{ddm} \ac{ofdm} waveform is explained in Sec.~\ref{sec:Communication_Setup_DDM}, and Sec.~\ref{sec:BER_performance} analyzes the \ac{ber} simulation results of this system. This work is concluded in Sec.~\ref{sec:Conclusio}.
\\
\\
\emph{Notation}: 
\\ 
Vectors and matrices are indicated by lower-case and upper-case bold face variables, respectively. The element of a matrix at its $l$th row and $k$th column is defined as $\left[ \m{A} \right]_{l,k}$, where the indices start with $0$. $\mathbb{R}$ and $\mathbb{C}$ represent the set of real and complex values, respectively. A superscript to $\mathbb{R}$ or $\mathbb{C}$ indicates the dimensions. Moreover, we use $\text{j}$ represents the imaginary unit, $(\cdot)^T$ denotes the transposition, $(\cdot)^H$ represents the conjugate transposition, $(\cdot)^*$ indicates complex conjugation. The identity matrix of size $n\times n$ is denoted as $\m{I}^{n}$, and a column vector of length $n$ with all elements equal to 1 is indicated by $\ve{1}^n$. The Hadamard product and Hadamard division are represented by $\odot$ and $\oslash$, respectively. 
\\
\\
\emph{Definitions}: 
\\ 
$\m{F}_N$ represents the \ac{dft} matrix of size $N \times N$ with $\left[ \m{F}_N \right]_{l,k} = \text{exp}\left( -\text{j} 2 \pi l k / N  \right)$ and $l,k = 0, \hdots, N-1$. 
The vector $\ve{d}_N(f) \in \mathbb{C}^{N}$ is defined as
\begin{equation}
	\ve{d}_N(f) = \begin{bmatrix}
	1  &
	\text{e}^{\text{j} 2 \pi f}  &
	\hdots  &
	 \text{e}^{\text{j} 2 \pi f(N-1)}
\end{bmatrix}^T,		\label{equ:MIMO_OFDM_005}
\end{equation}
with $f$ being a unitless place-holder variable. The matrix $\m{D}_N(f) \in \mathbb{C}^{N \times N}$ is a diagonal matrix defined as $\m{D}_N(f) = \text{diag} \left( \ve{d}_N(f) \right)$. 
Let $\m{W}_N = \mathrm{diag}\left( \ve{w}_N \right)\in \mathbb{R}^{N \times N}$ be a diagonal matrix containing the window function $\ve{w}_N \in \mathbb{R}^N$, then the windowed \ac{dft} of the complex-valued oscillation in $\ve{d}_N(f)$ yields \cite{Hakobyan_Inter_Carrier_Interference}
\begin{align}
	\ve{u}_N(f) &={} \m{F}_N\, \m{W}_N \, \ve{d}_N(f) \\
	 &={} \begin{bmatrix}
	\sum_{n=0}^{N-1} \left[ \ve{w}_N \right]_n \text{e}^{\text{j} 2 \pi (f - \frac{0}{N})n} \\
	\vdots  \\
	 \sum_{n=0}^{N-1} \left[ \ve{w}_N \right]_n \text{e}^{\text{j} 2 \pi (f - \frac{N-1}{N})n}
\end{bmatrix} \in \mathbb{C}^{N}.		\label{equ:MIMO_OFDM_008}
\end{align}
The vector $\ve{u}_N(f)$ contains a main peak whose position within the vector is determined by $f$. The remaining elements of $\ve{u}_N(f)$ contain either zeros or sidelobes of the main peak.

\section{Basics of OFDM Radar}  \label{sec:Basics_OFDM_Radar}

\begin{figure*}[!t]
\centering
\begin{tikzpicture}[scale=0.8, style=thick, rounded corners=1pt,inner sep=3.2pt,node distance=.8cm,every text node part/.style={align=center},
decoration = {snake,   
                    pre length=3pt,post length=7pt,
                    }]

\node[ minimum height = 1cm, minimum width = 1cm] (state0){\small Transmit data \\ \small $\m{S} \in \mathbb{C}^{N_\text{c} \times N_\text{sym}}$};
\node[draw,right=1cm of state0, minimum height = 1cm, minimum width = 1cm](state1){\small $\m{F}_{N_\text{c}}^{-1} \downarrow$};

\draw[->] (state0.east)  -- (state1.west);

\node[draw,right=1cm of state1, minimum height = 1cm, minimum width = 1cm](state2){\small Add CP};

\draw[->] (state1.east)  -- (state2.west);



\draw (state2.east)  -- ++(0.5,0) coordinate (Ant1);
\draw (Ant1)  -- ++(0,0.4) coordinate (IntAnt1);
\draw (IntAnt1)  -- ++(-0.2,0.3);
\draw (IntAnt1)  -- ++(0.2,0.3);

\node[draw,below=0.5cm of state2, minimum height = 1cm, minimum width = 1cm](state5){\small Remove \\ \small CP};

\draw (state5.east)  -- ++(0.5,0) coordinate (Ant2);
\draw (Ant2)  -- ++(0,0.4) coordinate (IntAnt2);
\draw (IntAnt2)  -- ++(-0.2,0.3);
\draw (IntAnt2)  -- ++(0.2,0.3);

\node[above right =-0.5cm and 2cm of state2.east, minimum height = 0cm, minimum width = 0cm, , rotate=90, anchor=north](Object){ Objects};

\draw [->, shorten <=0.2cm, shorten >=0.2cm] (IntAnt1) to (Object.north);
\draw [->, shorten <=0.2cm, shorten >=0.2cm] (Object.north) to (IntAnt2);

\node[left=0.4cm of state5](yarr1){};
\node[above=0.4cm of yarr1](yarr2){$\m{Y}_{\text{tf,ts}}$};
\draw [->, shorten <=0.01cm, shorten >=0.01cm, style=solid] (yarr1) to [out=80,in=260] (yarr2);


\node[draw,left=1cm of state5, minimum height = 1cm, minimum width = 1cm](state6){\small $\m{F}_{N_\text{c}} \downarrow$};

\draw[->] (state5.west)  -- (state6.east);

\node[draw,left=1cm of state6, minimum height = 1cm, minimum width = 1cm](state7){\small Element-wise \\ \small  division \\ \small with $\m{S}$ };

\draw[->] (state6.west)  -- (state7.east);

\node[draw,left=1cm of state7, minimum height = 1cm, minimum width = 1cm](state8){\small $\m{F}_{N_\text{c}}^{-1} \downarrow$};

\draw[->] (state7.west)  -- (state8.east);

\node[draw,left=1cm of state8, minimum height = 1cm, minimum width = 1cm](state9){\small $\m{F}_{N_\text{sym}} \rightarrow$};

\draw[->] (state8.west)  -- (state9.east);

\node[ left=1cm of state9, minimum height = 1cm, minimum width = 1cm] (state10){\small RDM of size \\ \small  $N_\text{c} \times N_\text{sym}$
};

\draw[->] (state9.west)  -- (state10.east);

\end{tikzpicture}
\caption{SISO OFDM radar signal processing chain, where the parallel-to-serial conversion, the ADC and DAC, and the analog front-end are not shown for simplicity. The arrows $\downarrow$ / $\rightarrow$ represent the dimension of a matrix on which the operation is applied. Figure taken from \cite{Lang_RDM_JP}.}
\label{fig:SISO_processing_steps}
\end{figure*}
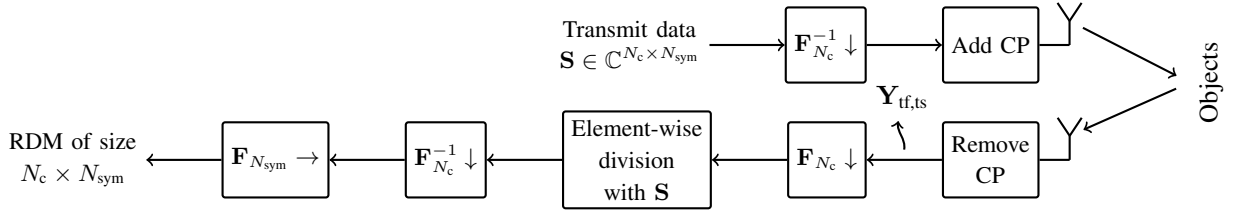

This section starts with a very brief textual explanation of the receiver signal processing chain of an \ac{ofdm}-based \ac{siso} radar system, which corresponds to that used for \ac{ddm} later in this work. After that, the signal model for the \ac{mimo} case is briefly described. The reader is referred to \cite{Sturm_A_novel_approach, Sturm_An_OFDM_System, Sturm_Performance_verification, Sit_Automotive_MIMO_OFDM,Braun_Parametrization, Hakobyan_Inter_Carrier_Interference, Lang_RDM_JP} for more details. 

\subsection{OFDM Waveform and Radar Signal Processing} 

\ac{ofdm} waveforms utilized in radar applications are typically based on the \ac{cp}-\ac{ofdm} waveform. \ac{cp}-\ac{ofdm} is widely adopted in wireless communications \cite{Van_Nee_OFDM} and the reader is thus expected to be familiar with it. Tab.~\ref{Tab:OFDM_paramters} lists important design parameters of the \ac{ofdm} waveform. 

\begin{table}[!t]
\renewcommand{\arraystretch}{1.3}
\caption{Parameter Definitions.}
\label{Tab:OFDM_paramters}
\centering
\begin{tabular}{|l|l|}
\hline
 Parameter & Symbol   \\
 \hline \hline
 Carrier frequency & $f_\text{c}$ \\
 \hline
 Bandwidth &  $B$ \\
 \hline
 Number of subcarriers & $N_\text{c}$ \\
  \hline
 ADC sampling time & $T_\text{s} =  1/B$ \\
 \hline
 Subcarrier spacing & $ \Delta f = B/ N_\text{c}$ \\
 \hline
 Length of an \ac{ofdm} symbol & $T = 1 / \Delta f$ \\
 \hline
 Length of the cyclic prefix  & $T_\text{cp}$ \\
 \hline
 Number of \ac{ofdm} symbols & $N_\text{sym}$ \\
 \hline
  Number of Rx antennas &  $N_\text{Rx}$ \\
 \hline
 Number of Tx antennas &  $N_\text{Tx}$ \\
 \hline
\end{tabular}
\end{table}

Fig.~\ref{fig:SISO_processing_steps} visualizes the principle \ac{siso} \ac{ofdm} radar signal processing chain. In this figure, matrix $\m{S} \in \mathbb{C}^{N_\text{c} \times N_\text{sym}}$ contains the complex-valued amplitudes for all $N_\text{c}$ subcarriers and for all $N_\text{sym}$ \ac{ofdm} symbols. The elements of $\m{S}$ are referred to as subcarrier symbols.
The \ac{ofdm} symbols are transformed into time domain, extended by a \ac{cp} to avoid \ac{isi} \cite{Van_Nee_OFDM}, and radiated by the Tx antenna. 
 
 The receiver senses signals reflected from objects and feeds them into the radar receiver signal processing chain, which consists of the following steps:
 \begin{enumerate}
 \item Removing the \ac{cp}.
 \item Applying a \ac{dft} to obtain the received frequency domain \ac{ofdm} symbols.
 \item Performing an element-wise division by the transmitted subcarrier symbols in $\m{S}$.
 \item Applying the so-called range \ac{idft} unveils the range information.
 \item Finally, the \ac{rvm} of size $N_\text{c} \times N_\text{sym}$ is obtained after the so-called Doppler \ac{dft}.
 \end{enumerate}
With $c_0$ denoting the speed of light, the final \ac{rvm} is determined by the values
\cite{Sturm_A_novel_approach, Lang_RDM_JP, Lang_Asilomar_2020}
\begin{align}
 \Delta r &= \frac{ c_0 }{2 B}    & \Delta v &= \frac{ c_0 }{2 f_\text{c} N_\text{sym} \left(T + T_{\text{cp}} \right)}   \label{equ:OFDM_001} \\
 r_\text{max} &= \Delta r  N_\text{c}   & v_\text{max} &= \pm \Delta v \frac{N_\text{sym}}{2}.  \label{equ:OFDM_004}
\end{align}

\subsection{MIMO Signal Model}

This section briefly recaps the complex baseband representation of the \ac{mimo} signal model from \cite{Lang_RDM_JP} since it is essential for deriving the \ac{ddm} method. A detailed derivation can be found in \cite{Lang_RDM_JP}, which itself is based on a \ac{siso} signal model derived in \cite{Hakobyan_Inter_Carrier_Interference}. The derivation of the signal model considers only a single Rx antenna, while a possible extension to multiple Rx antennas can be easily adapted. At first, some definitions and assumptions are introduced.


Let the matrices $\m{S}_k \in \mathbb{C}^{N_\text{c} \times N_\text{sym}}$ with $k=0, 1, \hdots , N_\text{Tx}-1$ contain the transmit subcarrier symbols for all Tx antennas. Each column of $\m{S}_k$ represents one frequency domain \ac{ofdm} symbol. The transmit signals in complex baseband can be derived by transforming these \ac{ofdm} symbols into time domain and extending them with a \ac{cp}.


The channel between transmitter and receiver assumes $N_{\text{path}}$ propagation paths between each of the $N_\text{Tx}$ Tx antennas and the Rx antenna with $r_{i,k}$ denoting the propagation distance for the $k$th Tx antenna along the $i$th path. $\tau_{i,k} = 2 r_{i,k} / c_0 $ is the corresponding round-trip delay time and can be normalized to $\bar{\tau}_{i,k} = \tau_{i,k} \Delta f$. $a_i\in \mathbb{C}$ models assumed constant amplitude and phase changes during propagation along the $i$th path \cite{Hakobyan_Inter_Carrier_Interference}. 

 At the receiver, the time domain \ac{adc} samples are stored in a matrix $\m{Y}_{\text{tf,ts}} \in \mathbb{C}^{N_\text{c} \times N_\text{sym}}$, where every column corresponds to a received \ac{ofdm} symbol in time domain and without the \ac{cp}. This matrix $\m{Y}_{\text{tf,ts}}$ is given by \cite{Lang_RDM_JP}
\begin{align}
	\m{Y}_{\text{tf,ts}} &={} \sum_{k=0}^{N_\text{Tx}-1} \sum_{i=0}^{N_\text{path}-1} \bar{a}_{i,k} \m{D}_{N_\text{c}}\left( \frac{ \bar{f}_{\text{D}_i}}{N_\text{c}} \right) \m{F}_{N_\text{c}}^{-1} \m{D}_{N_\text{c}}^*(\bar{\tau}_{i,k}) \nonumber \\
	& \hspace{4mm} \cdot \m{S}_k \m{D}_{N_\text{sym}}( \bar{f}_{\text{D}_i} \alpha ).\label{equ:MIMO_OFDM_011_2}
\end{align}
where additive measurement noise is neglected, and where 'tf' and 'ts' indicate the fast time and the slow time over the vertical and horizontal matrix dimension, respectively. Additionally, we used $\bar{a}_{i,k}  = a_i \, \text{exp}(-\text{j} 2 \pi f_\text{c} \tau_{i,k})\in \mathbb{C}$ and $\alpha = (T + T_\text{cp})/T \in \mathbb{R}$. 
The Doppler shift along the $i$th path caused by a relative velocity $v_i$ is defined as $f_{\text{D}_{i}} = -2 v_i f_\text{c} / c_0$, and it is normalized to $\bar{f}_{\text{D}_i} = f_{\text{D}_i} / \Delta f$.
The implications of the Doppler shift are considered in form of a \ac{cpe} and \ac{ici}  affecting the received \ac{ofdm} symbols. The \ac{cpe} and \ac{ici} are represented in \eqref{equ:MIMO_OFDM_011_2} in form of $\m{D}_{N_\text{sym}}( \bar{f}_{\text{D}_i} \alpha )$ and $\m{D}_{N_\text{c}}\left( \frac{ \bar{f}_{\text{D}_i}}{N_\text{c}} \right)$, respectively.

For reasons of compactness of the subsequent mathematical derivations, we approximate the \ac{ici} term as $\m{D}_{N_\text{c}}\left( \frac{ \bar{f}_{\text{D}_i}}{N_\text{c}} \right) \approx \m{I}^{N_\text{c}}$, which is valid for moderate $v_i$ and sufficiently large $\Delta f$, leading to
\begin{align}
	\m{Y}_{\text{tf,ts}} &={} \sum_{k=0}^{N_\text{Tx}-1} \sum_{i=0}^{N_\text{path}-1} \bar{a}_{i,k} \m{F}_{N_\text{c}}^{-1} \m{D}_{N_\text{c}}^*(\bar{\tau}_{i,k})\m{S}_k \m{D}_{N_\text{sym}}( \bar{f}_{\text{D}_i} \alpha ).\label{equ:MIMO_OFDM_011}
\end{align}
However, \ac{ici} as well as additive measurement noise are fully considered for all simulations in this work unless clearly stated otherwise.


In case several Rx antennas are considered, a similar matrix as in \eqref{equ:MIMO_OFDM_011} can be constructed for every Rx antenna with appropriately modified parameters $\bar{a}_{i,k}$ and $\bar{\tau}_{i,k}$ \cite{Lang_RDM_JP}. 


\section{Doppler-Division Multiplexing} \label{sec:Proposed_MIMO}


The key aspect of \ac{ddm} is a modification of the individual transmit signals for every Tx antenna such that the corresponding receive signals are shifted along the velocity axis in the \ac{rvm}. These modified transmit signals as well as the resulting \ac{rvm} are derived in the following.


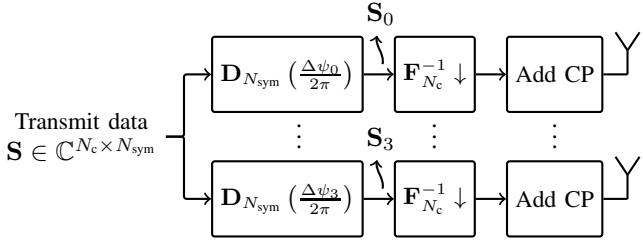
\begin{figure}[!t]
\centering
\begin{tikzpicture}[scale=0.8, style=thick, rounded corners=1pt,inner sep=3.2pt,node distance=.8cm,every text node part/.style={align=center}]

\node[ minimum height = 1cm, minimum width = 1cm] (state0){\small Transmit data \\ $\m{S} \in \mathbb{C}^{N_\text{c} \times N_\text{sym}}$};

\node[above right =-0.8cm and 1.6cm of state0, minimum height = 0cm, minimum width = 0cm]{ $\vdots $};
\node[above right =-0.8cm and 3.4cm of state0, minimum height = 0cm, minimum width = 0cm]{ $\vdots $};
\node[above right =-0.8cm and 5.0cm of state0, minimum height = 0cm, minimum width = 0cm]{ $\vdots $};

\node[draw, above right =-0.2cm and 0.6cm of state0, minimum height = 1cm, minimum width = 1cm](state1u){\small $ \m{D}_{N_\text{sym}} \left( \frac{\Delta \psi_0}{2 \pi} \right)  $};

\node[right=0.1cm of state1u](yarr1){};
\node[above=0.4cm of yarr1](yarr2){$\m{S}_{0}$};
\draw [->, shorten <=0.01cm, shorten >=0.01cm, style=solid] (yarr1) to [out=80,in=260] (yarr2);

\node[draw, below right =-0.2cm and 0.6cm of state0, minimum height = 1cm, minimum width = 1cm](state1d){\small $ \m{D}_{N_\text{sym}} \left( \frac{\Delta \psi_3}{2 \pi} \right)  $};

\node[right=0.1cm of state1d](yarr1a){};
\node[above=0.4cm of yarr1a](yarr2a){$\m{S}_{3}$};
\draw [->, shorten <=0.01cm, shorten >=0.01cm, style=solid] (yarr1a) to [out=80,in=260] (yarr2a);

\draw[->] (state0.east)  -- ++(0.3cm,0) -- ++(0,1.035cm) -- (state1u.west);
\draw[->] (state0.east)  -- ++(0.3cm,0) -- ++(0,-1.035cm) -- (state1d.west);

\node[draw,right=0.4cm of state1u, minimum height = 1cm, minimum width = 1cm](state2u){\small $\m{F}_{N_\text{c}}^{-1} \downarrow$};
\node[draw,right=0.4cm of state1d, minimum height = 1cm, minimum width = 1cm](state2d){\small $\m{F}_{N_\text{c}}^{-1} \downarrow$};

\draw[->] (state1u.east)  -- (state2u.west);
\draw[->] (state1d.east)  -- (state2d.west);

\node[draw,right=0.4cm of state2u, minimum height = 1cm, minimum width = 1cm](state3u){\small Add CP};
\node[draw,right=0.4cm of state2d, minimum height = 1cm, minimum width = 1cm](state3d){\small Add CP};

\draw[->] (state2u.east)  -- (state3u.west);
\draw[->] (state2d.east)  -- (state3d.west);

%

\draw (state3u.east)  -- ++(0.4,0) coordinate (Ant1u);
\draw (Ant1u)  -- ++(0,0.4) coordinate (IntAnt1u);
\draw (IntAnt1u)  -- ++(-0.2,0.3);
\draw (IntAnt1u)  -- ++(0.2,0.3);

\draw (state3d.east)  -- ++(0.4,0) coordinate (Ant1d);
\draw (Ant1d)  -- ++(0,0.4) coordinate (IntAnt1d);
\draw (IntAnt1d)  -- ++(-0.2,0.3);
\draw (IntAnt1d)  -- ++(0.2,0.3);

\end{tikzpicture}
\caption{DDM signal processing chain in the transmitter for the case of $N_\text{Tx}=4$. The the parallel-to-serial conversion, the DAC, and the analog front-end are not shown for simplicity.}
\label{fig:DDM_Transmitter}
\end{figure}

The derivation begins with the \ac{mimo} signal model in \eqref{equ:MIMO_OFDM_011}. A shift along the velocity axis in the \ac{rvm} is implemented by applying a phase shift from \ac{ofdm} symbol to \ac{ofdm} symbol. This phase shift is denoted as $\Delta \psi_k$ for $0 \leq k < N_\text{Tx}$. For instance, the first \ac{ofdm} symbol of the $k$th Tx antenna remains unchanged, while the follow-up \ac{ofdm} symbols are rotated in phase by $\Delta \psi_k$, $2\Delta \psi_k$, and so on.

Applying this phase shift in the transmitter is achieved by choosing $\m{S}_k$ in \eqref{equ:MIMO_OFDM_011} to be
\begin{align}
	\m{S}_k = \m{S}  \m{D}_{N_\text{sym}} \left( \frac{\Delta \psi_k}{2 \pi} \right), \label{equ:MIMO_OFDM_010}
\end{align}
cf. Fig.~\ref{fig:DDM_Transmitter}. Note that $\m{S}$, and thus the payload, is the same for all Tx antennas. Only the modulation by $\m{D}_{N_\text{sym}} \left( \frac{\Delta \psi_k}{2 \pi} \right)$ makes them distinguishable among each other. 

The time domain \ac{ofdm} symbols at the $k$th Tx antenna in complex baseband are obtained by transforming the \ac{ofdm} symbols in \eqref{equ:MIMO_OFDM_010} into time domain and extending them by a \ac{cp}.


The receive signal in complex baseband representation and after removing the \ac{cp} is transformed into frequency domain by applying the \ac{dft} on the columns of $\m{Y}_{\text{tf,ts}}$, yielding
\begin{align}
	& \m{Y}_{\text{f,ts}} ={}  \m{F}_{N_\text{c}} \m{Y}_{\text{tf,ts}} \\
	&={} \sum_{k=0}^{N_\text{Tx}-1} \sum_{i=0}^{N_\text{path}-1} \bar{a}_{i,k} \m{D}_{N_\text{c}}^*(\bar{\tau}_{i,k})  \m{S} \m{D}_{N_\text{sym}} \left( \frac{\Delta \psi_k}{2 \pi} + \bar{f}_{\text{D}_i} \alpha \right). \label{equ:MIMO_OFDM_012}
\end{align}
The subscript 'f' indicates that the columns of $\m{Y}_{\text{f,ts}}$ represent the frequency domain \cite{Hakobyan_Inter_Carrier_Interference}. The second step is the Hadamard (element-wise) division with $\m{S}$, yielding  \cite{Lang_RDM_JP}
\begin{align}
	&\m{Z}_{\text{f,ts}} ={}  \m{Y}_{\text{f,ts}} \oslash \m{S} \\
	&={} \sum_{k=0}^{N_\text{Tx}-1} \sum_{i=0}^{N_\text{path}-1} \bar{a}_{i,k} \m{D}_{N_\text{c}}^*(\bar{\tau}_{i,k}) \ve{1}^{N_\text{c}} (\ve{1}^{N_\text{sym}})^T  \nonumber \\
	& \hspace{4mm} \cdot  \m{D}_{N_\text{sym}} \left( \frac{\Delta \psi_k}{2 \pi} + \bar{f}_{\text{D}_i} \alpha \right) \\
	&={} \sum_{k=0}^{N_\text{Tx}-1} \sum_{i=0}^{N_\text{path}-1} \bar{a}_{i,k} \ve{d}_{N_\text{c}}^*(\bar{\tau}_{i,k})  \ve{d}_{N_\text{sym}}^T \left( \frac{\Delta \psi_k}{2 \pi} + \bar{f}_{\text{D}_i} \alpha \right). \label{equ:MIMO_OFDM_013}
\end{align}


The third step is applying the windowed range \ac{idft} on the columns of $\m{Z}_{\text{f,ts}}$ according to \cite{Hakobyan_Inter_Carrier_Interference}
\begin{align}
	&\m{Z}_{\text{r,ts}} ={} \m{F}_{N_\text{c}}^{-1} \m{W}_{N_\text{c}} \m{Z}_{\text{f,ts}} \\
	&={} \sum_{k=0}^{N_\text{Tx}-1} \sum_{i=0}^{N_\text{path}-1} \bar{a}_{i,k} \ve{u}_{N_\text{c}}^*(\bar{\tau}_{i,k})  \ve{d}_{N_\text{sym}}^T \left( \frac{\Delta \psi_k}{2 \pi} + \bar{f}_{\text{D}_i} \alpha \right). \label{equ:MIMO_OFDM_014}
\end{align}
The fourth processing step is applying the windowed Doppler \ac{dft} on the rows of $\m{Z}_{\text{r,ts}}$, which yields
\begin{align}
	&\m{Z}_{\text{r,v}} ={} \m{Z}_{\text{r,ts}} \m{W}_{N_\text{sym}} \m{F}_{N_\text{sym}} \\
	&={}\sum_{k=0}^{N_\text{Tx}-1} \sum_{i=0}^{N_\text{path}-1} \bar{a}_{i,k} \ve{u}_{N_\text{c}}^*(\bar{\tau}_{i,k}) \ve{u}_{N_\text{sym}}^T \left( \frac{\Delta \psi_k}{2 \pi} + \bar{f}_{\text{D}_i} \alpha \right).\label{equ:MIMO_OFDM_015}
\end{align}
This result represents the final \ac{rvm} containing $N_\text{Tx}$ peaks for every path $i$. These peaks are located at the same range but at different velocities.

\section{Discussion of DDM in Context of Radar Sensing} \label{sec.Properties_DDM}


\subsection{Choice of $\Delta \psi_k$} \label{sec.Choice_of_phi}
The shift of the signal components along the velocity axis is determined by $\Delta \psi_k$. It is recommended to choose $\Delta \psi_k=2 \pi \frac{p}{N_\text{sym}}$ for any $p \in \mathbb{Z}$, which circularly shifts the corresponding signal components by $p$ velocity bins without changing the magnitude or phase values. This statement can be proven by a straightforward modification of a related proof in \cite[Appendix A]{Lang_RDM_JP}. Since no distortions of the magnitude or phase values are induced, a utilization of the them for \ac{dbf} \cite{saponara2017radar, patole2017automotive, Gerstmair1ADAS} is easily possible. 

Furthermore, for $N_\text{Tx} = 4$, it is easy to prove that the choice $\Delta \psi_k = \{ -\frac{3 \pi}{4}, -\frac{1 \pi}{4}, \frac{1 \pi}{4}, \frac{3 \pi}{4}\}$  separates the \ac{rvm} in 4 equally sized areas, as sketched in Fig.~\ref{fig_DDM_DDM}, in which the bin representing zero relative velocity is located at the center of each area. Thus, it will be the primary choice for $\Delta \psi_k$ in this work. 

We note that employing a \ac{dft} precoding matrix of size $N_\text{Tx} \times N_\text{Tx}$ as utilized in \cite{suh2021time} may produce the same phase shift values as used in this work, however, it offers less freedom in choosing the shift along the velocity axis.

\subsection{Maximum Unambiguous Range and Velocity}
\ac{ddm} provides the same unambiguous range $r_\text{max}$ as for a \ac{siso} \ac{ofdm} radar system in \eqref{equ:OFDM_004}. However, when dividing the velocity axis into $N_\text{Tx}$ equally sized areas as discussed in Sec.~\ref{sec.Choice_of_phi}, the maximum unambiguous velocity $v_\text{max}$ is decreased for \ac{ddm} by a factor of $N_\text{Tx}$ compared to the  \ac{siso} case (cf. Fig.~\ref{fig_DDM_DDM}). As a consequence, the number of Tx antennas supported by \ac{ddm} without further measures can be increased as long as no object violates the reduced maximum unambiguous velocity.

\subsection{Beampattern}
The beampattern is an important performance criterion for phased arrays and specifies the average signal power transmitted towards a certain direction. 
Simulations confirmed that the beampattern for \ac{ddm} is almost uniform for practical values of $N_\text{sym}$. 

\subsection{Computational Complexity}

The computational complexity required for adding the phase shift from \ac{ofdm} symbol to \ac{ofdm} symbol  depends on $\Delta \psi_k$. In the worst case, $N_\text{c} N_\text{sym}$  phase rotations are required to apply the phase shift on every subcarrier in every \ac{ofdm} symbol. For the special choice of $\Delta \psi_k = \{ -\frac{3 \pi}{4}, -\frac{1 \pi}{4}, \frac{1 \pi}{4}, \frac{3 \pi}{4}\} $, the phase shifts are multiples of $\frac{\pi}{4}$ and thus computational complexity may be much lower. More specifically, if the symbol alphabet is symmetric with respect to rotations of $\frac{\pi}{4}$, applying $\Delta \psi_k$ can simply be implemented by modifying the so-called mapper with a time-dependent mapping function. For symbol alphabets without this symmetry, an extension of the alphabet might be considered. 


\subsection{Comparison with ESI} 

\ac{esi} has the following similarities and differences compared to \ac{ddm}. 

\emph{Processing gain}: Since \ac{ddm} activates every subcarrier on all Tx antennas, the processing gain follows as $G_\text{p} = N_\text{sym} N_\text{c}$. In contrast to that, \ac{esi} activates only every $N_\text{Tx}$th subcarrier per Tx antenna, leading to a reduced processing gain of $G_\text{p} = N_\text{sym} N_\text{c} / N_\text{Tx}$. 

\emph{Average power per active subcarrier}: Let \ac{esi} and \ac{ddm} have the same average transmit power in order to provide a fair comparison. Then, as a consequence of the fewer activated subcarriers for \ac{esi}, the average power per active subcarrier is $N_\text{Tx}$ times larger for \ac{esi} compared to \ac{ddm}. 

\emph{SNR in the RDM}: Due to the same argumentation as provided in \cite{Lang_RDM_JP}, the reduced power per active subcarrier and the increased processing gain cancel each other out. This results in approximately the same \ac{snr} in the \ac{rvm}, and as a direct consequence, also the same \ac{snr} at the output of the \ac{dbf}, for \ac{ddm} as for \ac{esi}. 

\emph{Maximum unambiguous range and velocity}: 
As a consequence of the reduced number of active subcarriers per Tx antenna, \ac{esi} reduces $r_\text{max}$ by a factor of $N_\text{Tx}$ compared to a \ac{siso} \ac{ofdm} system. In contrast to that, \ac{ddm} offers the same $r_\text{max}$ as a \ac{siso} \ac{ofdm} system, but it reduces the unambiguous maximum velocity $v_\text{max}$ by a factor of $N_\text{Tx}$.



\subsection{Comparison with RDMult}

\ac{rdm} in \cite{Lang_RDM_JP} shifts signal components along the range axis, and thus, reduces $r_\text{max}$ by a factor of $N_\text{Tx}$ compared to the \ac{siso} case. \ac{ddm} shifts signal components along the velocity axis in the \ac{rvm}, which entails a reduction of $v_\text{max}$ by a factor of $N_\text{Tx}$ compared the \ac{siso} case. Despite this difference, both multiplexing methods feature the same average power per active subcarrier, the same processing gain, and the same \ac{snr} in the \ac{rvm}.

\section{Communication System based on DDM} \label{sec:Communication_Setup_DDM}

The \ac{ddm} method generates transmit signals designed for multiplexing purposes in radar sensing applications. The effects of the transmit signal design due to \ac{ddm} on the communication task are investigated in this section. Based on these investigations, a communication system specifically designed for \ac{ddm} is proposed. In the following, the terms 'receiver' and 'Rx antenna' do not refer to the radar receiver but to the communication receiver.

This section begins with a discussion on different ways to represent the channel for \ac{ddm}, followed by deriving the so-called 'effective' channel. This effective channel is a time-varying \ac{siso} channel that sufficiently describes the communication link between the transmitter and the communication receiver. After that, estimation methods for the effective channel and the transmit data are proposed. 
For these investigations, the waveform and system parameters are chosen according to Tab.~\ref{Tab:com_sys_paramters}. The communication receiver employs only a single Rx antenna, which will turn out to be sufficient for enabling communications. An extension to multiple Rx antennas is straightforward at the cost of additional hardware and an increased power consumption for the communication receiver. 

\begin{table}[!t]
\renewcommand{\arraystretch}{1.3}
\caption{Waveform and system parameters.}
\label{Tab:com_sys_paramters}
\centering
\begin{tabular}{|l|r|}
\hline
 Parameter & Value  \\
 \hline \hline
 Carrier frequency $f_\text{c}$ & $77 \, \text{GHz}$ \\
 \hline
 Bandwidth $B$ & $1\, \text{GHz}$ \\
 \hline
 Number of subcarriers $N_\text{c}$ & $1024$ \\
 \hline
 \ac{adc} sampling time $T_\text{s}$  &  $1\, \text{ns}$ \\
 \hline
 Subcarrier spacing $\Delta f$ & $976.5 \, \text{MHz}$ \\
 \hline
 
 Length of an OFDM symbol $T$ & $1.024\,$\textmu$\text{s}$ \\
 \hline
 Length of the cyclic prefix $T_\text{cp}$ & $1\,$\textmu$\text{s}$ \\
 \hline
 Number of \ac{ofdm} symbols $N_\text{sym}$ & $512$ \\
 \hline
  Number of Rx antennas $N_\text{Rx}$ & $1$ \\
 \hline
 Number of Tx antennas $N_\text{Tx}$ & $4$ \\
 \hline
  Symbol alphabet & QPSK \\
 \hline
 Phase shift $\Delta \psi_k$ &  $\{ -\frac{3 \pi}{4}, -\frac{1 \pi}{4}, \frac{1 \pi}{4}, \frac{3 \pi}{4}\}$ \\
 \hline
\end{tabular}
\end{table}

\subsection{Channel Model} 

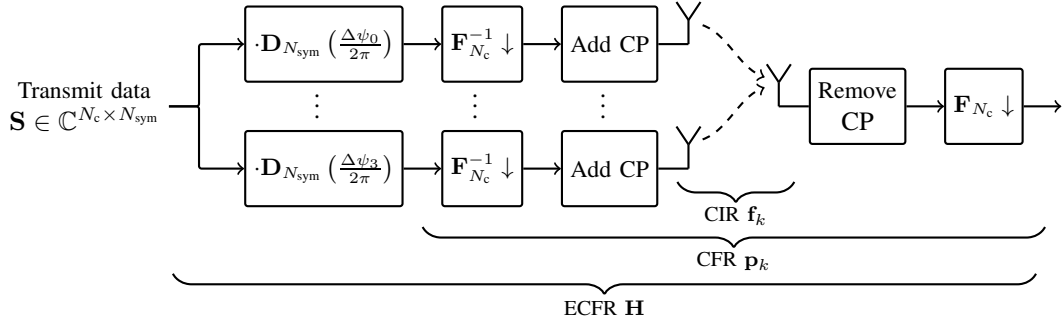
\begin{figure*}[!t]
\centering
\begin{tikzpicture}[scale=0.8, style=thick, rounded corners=1pt,inner sep=3.2pt,node distance=.8cm,every text node part/.style={align=center}]

\node[ minimum height = 1cm, minimum width = 1cm] (state0){\small Transmit data \\ $\m{S} \in \mathbb{C}^{N_\text{c} \times N_\text{sym}}$};

\node[above right =-0.8cm and 1.8cm of state0, minimum height = 0cm, minimum width = 0cm]{ $\vdots $};
\node[above right =-0.8cm and 3.9cm of state0, minimum height = 0cm, minimum width = 0cm]{ $\vdots $};
\node[above right =-0.8cm and 5.6cm of state0, minimum height = 0cm, minimum width = 0cm]{ $\vdots $};

\node[draw, above right =-0.2cm and 1cm of state0, minimum height = 1cm, minimum width = 1cm](state1u){\small $ \cdot\m{D}_{N_\text{sym}} \left( \frac{\Delta \psi_0}{2 \pi} \right)  $};

\node[draw, below right =-0.2cm and 1cm of state0, minimum height = 1cm, minimum width = 1cm](state1d){\small $ \cdot \m{D}_{N_\text{sym}} \left( \frac{\Delta \psi_3}{2 \pi} \right)  $};

\draw[->] (state0.east)  -- ++(0.5cm,0) -- ++(0,1.035cm) -- (state1u.west);
\draw[->] (state0.east)  -- ++(0.5cm,0) -- ++(0,-1.035cm) -- (state1d.west);

\node[draw,right=0.5cm of state1u, minimum height = 1cm, minimum width = 1cm](state2u){\small $\m{F}_{N_\text{c}}^{-1} \downarrow$};
\node[draw,right=0.5cm of state1d, minimum height = 1cm, minimum width = 1cm](state2d){\small $\m{F}_{N_\text{c}}^{-1} \downarrow$};

\draw[->] (state1u.east)  -- (state2u.west);
\draw[->] (state1d.east)  -- (state2d.west);

\node[draw,right=0.5cm of state2u, minimum height = 1cm, minimum width = 1cm](state3u){\small Add CP};
\node[draw,right=0.5cm of state2d, minimum height = 1cm, minimum width = 1cm](state3d){\small Add CP};

\draw[->] (state2u.east)  -- (state3u.west);
\draw[->] (state2d.east)  -- (state3d.west);

%

\draw (state3u.east)  -- ++(0.5,0) coordinate (Ant1u);
\draw (Ant1u)  -- ++(0,0.4) coordinate (IntAnt1u);
\draw (IntAnt1u)  -- ++(-0.2,0.3);
\draw (IntAnt1u)  -- ++(0.2,0.3);

\draw (state3d.east)  -- ++(0.5,0) coordinate (Ant1d);
\draw (Ant1d)  -- ++(0,0.4) coordinate (IntAnt1d);
\draw (IntAnt1d)  -- ++(-0.2,0.3);
\draw (IntAnt1d)  -- ++(0.2,0.3);


\node[draw,below right =-0.2cm and 2cm of state3u, minimum height = 1cm, minimum width = 1cm](state6){\small Remove \\ CP};

\draw (state6.west)  -- ++(-0.5,0) coordinate (Ant2);
\draw (Ant2)  -- ++(0,0.4) coordinate (IntAnt2);
\draw (IntAnt2)  -- ++(-0.2,0.3);
\draw (IntAnt2)  -- ++(0.2,0.3);

\draw [->, shorten <=0.2cm, shorten >=0.2cm, style=dashed] (IntAnt1u) to [out=-20,in=160] (IntAnt2);
\draw [->, shorten <=0.2cm, shorten >=0.2cm, style=dashed] (IntAnt1d) to [out=20,in=200] (IntAnt2);


\node[draw,right=0.5cm of state6, minimum height = 1cm, minimum width = 1cm](state7){\small $\m{F}_{N_\text{c}} \downarrow$};

\draw[->] (state6.east)  -- (state7.west);

\node[right=0.5cm of state7, minimum height = 0cm, minimum width = 0cm](state8){};

\draw[->] (state7.east)  -- (state8.west);

\node[below right =0.05cm and 0.0cm of state3d.east, minimum height = 0cm, minimum width = 0cm](cb1){};

\draw [decorate,decoration={brace,mirror, amplitude=8pt},xshift=0pt,yshift=0pt]
(cb1) -- ++(2.1,0.0) node [black,midway,yshift=-0.45cm] 
{\footnotesize CIR $\ve{f}_k$};

\node[below right =0.65cm and 0.0cm of state1d.east, minimum height = 0cm, minimum width = 0cm](cb1){};

\draw [decorate,decoration={brace,mirror, amplitude=8pt},xshift=0pt,yshift=0pt]
(cb1) -- ++(10.5,0.0) node [black,midway,yshift=-0.45cm] 
{\footnotesize CFR $\ve{p}_k$};

\node[below right =2.10cm and -0.2cm of state0.east, minimum height = 0cm, minimum width = 0cm](cb1){};

\draw [decorate,decoration={brace,mirror, amplitude=8pt},xshift=0pt,yshift=0pt]
(cb1) -- ++(14.5,0.0) node [black,midway,yshift=-0.45cm] 
{\footnotesize ECFR  $\m{H}$};

%
%
%
%

\end{tikzpicture}
\caption{Transmitter and receiver processing chains including the CIR, the CFR, and the ECFR with their covered processing blocks. The parallel-to-serial conversions, the DACs, the ADC, and the analog front-ends are not shown for simplicity. The figure is based on a similar figure in \cite{Lang_RDM_JP}.}
\label{fig:Comm_setup_ECIR}
\end{figure*}

Fig.~\ref{fig:Comm_setup_ECIR} shows the processing blocks for \ac{ddm} in the transmitter, the $N_\text{Tx}$ \acp{cir} from each Tx antenna to the Rx antenna, and the first two receiver processing blocks. On basis of that, we will introduce different channel representations


As for \ac{rdm} in \cite{Lang_RDM_JP}, the channel for \ac{ddm} can be represented in three possible ways. 

\subsubsection{CIR Representation} 

A straightforward way of describing the channel is to utilize the $N_\text{Tx}$ individual \acp{cir}. The model for these \acp{cir} \cite{Rappaport_SISO_Channel, Rappaport_SISO_Channel_JP, Code_MIMO_Channels} and the model parametrization are equal to that employed in \cite{Lang_RDM_JP}, such that a detailed description is omitted in this work. These \acp{cir} are denoted as $\ve{f}_k \in \mathbb{C}^{N_\ve{f}}$ for $0 \leq k < N_\text{Tx}$ and their assumed length is $N_\ve{f} = 256$. 

\subsubsection{CFR Representation} 

Another way is to utilize the \acp{cfr}, which describe the $N_\text{Tx}$ individual \acp{cir} in frequency domain. 
These \acp{cfr} are denoted as  $\ve{p}_k \in \mathbb{C}^{N_\text{c}}$ and are given by
\begin{align}
 \ve{p}_k = \m{F}_{N_\text{c}} \m{B}_{\text{zp}} \ve{f}_k, \label{equ:Comm_OFDM_ECIR_002o}
\end{align}
where the matrix $\m{B}_{\text{zp}}\in \mathbb{C}^{N_\text{c} \times N_\ve{f}}$ zero-pads the \acp{cir} $\ve{f}_k$ to a length of $N_\text{c}$.

\subsubsection{ECFR Representation} 

Here, we exploit the fact that all antennas transmit the same subcarrier symbols $\m{S}$ up to the deterministic phase shift $\Delta \psi_k$. Thus, the third way of describing the channel covers all shown processing blocks in Fig.~\ref{fig:Comm_setup_ECIR}. This channel representation is referred to as \ac{ecfr}, which can be modeled as a \ac{siso} channel despite the fact that several Tx antennas are involved. 

The \ac{ecfr} is mathematically described in the following. 
As depicted in Fig.~\ref{fig:Comm_setup_ECIR}, the \ac{ecfr} covers the \acp{cfr} and the diagonal matrices $\m{D}_{N_\text{sym}} \left( \frac{\Delta \psi_k}{2 \pi} \right)$. These diagonal matrices apply a phase shift from \ac{ofdm} symbol to \ac{ofdm} symbol. Thus, the \ac{ecfr} will change from \ac{ofdm} symbol to \ac{ofdm} symbol. Let $ \m{H} \in \mathbb{C}^{N_\text{c} \times N_\text{sym}}$ denote the \ac{ecfr} for all $N_\text{sym}$ \ac{ofdm} symbols, then, $ \m{H}$ is given by
\begin{align}
 \m{H} = \sum_{k=0}^{N_\text{Tx}-1} \m{Q}_k, \label{equ:Comm_OFDM_ECIR_002}
\end{align}
where the matrices $\m{Q}_k \in \mathbb{C}^{N_\text{c} \times N_\text{sym}}$ are given as
\begin{align}
 \m{Q}_k &={}  \ve{p}_k \left( \ve{1}^{N_\text{sym}} \right)^T \m{D}_{N_\text{sym}} \left( \frac{\Delta \psi_k}{2 \pi} \right) \\
 &={}  \ve{p}_k  \ve{d}_{N_\text{sym}} \left( \frac{\Delta \psi_k}{2 \pi} \right)^T. \label{equ:Comm_OFDM_ECIR_002qw}
\end{align}
Each matrix $\m{Q}_k$ for $0 \leq k < N_\text{Tx}$ represents one signal path in Fig.~\ref{fig:Comm_setup_ECIR}.

    
\subsubsection{Properties of the ECFR} \label{sec:Discussion_ECFR}
    
Note that the $N_\text{Tx}$ different terms used to construct the \ac{ecfr} in \eqref{equ:Comm_OFDM_ECIR_002} may interfere constructively or destructively. On top of that, this constructive/destructive interference turns out to be heavily time-varying. 

The time-dependency of the \ac{ecfr} can be well demonstrated for \ac{awgn} channels where  $\m{Q}_k$ in \eqref{equ:Comm_OFDM_ECIR_002qw} reduces to 
\begin{align}
 \m{Q}_k =  \ve{1}^{N_\text{c}} \ve{d}_{N_\text{sym}} \left( \frac{\Delta \psi_k}{2 \pi} \right)^T.   \label{equ:Comm_OFDM_ECIR_002ay}
\end{align}
According to \eqref{equ:Comm_OFDM_ECIR_002ay}, all subcarriers experience the same effects. It is thus sufficient to inspect a single subcarrier, e.g, the first one. 

This subcarrier is represented by the first row of $\m{Q}_k$ denoted as $\left[\m{Q}_k \right]_{0,\mu}$, where the \ac{ofdm} symbols are indexed with $0 \leq \mu < N_\text{sym}$. The magnitude and phase values for these elements are exemplarily sketched in form of arrows in the complex plane in Tab.~\ref{table_phase_rotations} for the first 9 \ac{ofdm} symbols $0 \leq \mu < 9$ and for the choice of $\Delta \psi_k = \{ \frac{1 \pi}{4}, \frac{3 \pi}{4}, \frac{5 \pi}{4}, \frac{7 \pi}{4}\}$. According to \eqref{equ:Comm_OFDM_ECIR_002}, the sum of the elements $\left[\m{Q}_k \right]_{0,\mu}$ for $0 \leq k < N_\text{Tx}$ yields the first subcarrier of the \ac{ecfr} $\left[\m{H} \right]_{0,\mu}$ for the $\mu$th \ac{ofdm} symbol, which is also sketched in Tab.~\ref{table_phase_rotations}. 
One can see, that for $\mu = 0$ all components add up constructively. Hence, the \ac{ecfr} can be considered to be good and the received signal power will be maximized for this \ac{ofdm} symbol. The next three \ac{ofdm} symbols observe destructive interference leading to full cancellation of the signal components, such that the received signal power is zero. For the \ac{ofdm} symbols $\mu=4, \hdots 7$ the arrows point in the opposite direction than for $\mu=0, \hdots 3$, respectively. This pattern repeats for $\mu \geq 8$. 
For the \ac{awgn} case, we notice that constructive/destructive interference is observed with a period of 8 \ac{ofdm} symbols. Within this period, the \ac{ecfr} for the first 4 \ac{ofdm} symbols equals that for the subsequent 4 \ac{ofdm} symbols when inverting all signs.

\begin{table}[!t]
\renewcommand{\arraystretch}{1.20}
\caption{Schematic visualization of the elements of $\left[\m{Q}_k \right]_{0,\mu}$ for $0 \leq k < N_\text{Tx}$ and for $0 \leq \mu < 9$ in case of \ac{awgn} channels.}
\label{table_phase_rotations}
\centering
\begin{tabular}{|c|c|c|c|c|c|c|c|c|c|}
\hline
 $\mu$&  0 & 1 & 2 & 3 & 4 & 5 & 6 & 7 & 8 \\
 \hline \hline
 $\left[\m{Q}_0 \right]_{0,\mu}$ &  $\rightarrow$ &  $\myswarrow$ &  $\uparrow$ &  $\mysearrow$ &  $\leftarrow$ & $\mynearrow$ & $\downarrow$ &  $\mynwarrow$ &  $\rightarrow$ \\
 \hline
  $\left[\m{Q}_1 \right]_{0,\mu}$ &  $\rightarrow$ &  $\mysearrow$ &  $\downarrow$ &  $\myswarrow$ &  $\leftarrow$ & $\mynwarrow$ & $\uparrow$ & $\mynearrow$ &  $\rightarrow$ \\
 \hline
  $\left[\m{Q}_2 \right]_{0,\mu}$ &   $\rightarrow$ &  $\mynearrow$ &  $\uparrow$ &  $\mynwarrow$ &  $\leftarrow$ & $\myswarrow$ & $\downarrow$ & $\mysearrow$ &  $\rightarrow$ \\
 \hline
  $\left[\m{Q}_3 \right]_{0,\mu}$ &  $\rightarrow$ &  $\mynwarrow$ &  $\downarrow$ &  $\mynearrow$ &  $\leftarrow$ & $\mysearrow$ & $\uparrow$ &  $\myswarrow$ &  $\rightarrow$ \\
 \hline \hline
 $\left[\m{H} \right]_{0,\mu}$ &  $\Largerightarrow$ &  0 &  0 &  0 &  $\Largeleftarrow$ & 0 & 0 &  0 &  $\Largerightarrow$ \\
 \hline
\end{tabular}
\end{table}

\begin{figure}[!t]
\begin{center}
\begin{tikzpicture}
\begin{axis}[compat=newest, 
width=0.8\columnwidth, height = .5\columnwidth, xlabel={OFDM symbol index $\mu$}, 
ylabel style={align=center}, 
ylabel style={text width=3.4cm},
ylabel={Normalized power of $\left[\m{H} \right]_{0,\mu}$ (dB)}, 
legend pos=north east, 
legend cell align=left,
legend columns=2, 
        legend style={
            /tikz/column 2/.style={
                column sep=5pt,
            },
        font=\small},
xmin = 0,
xmax = 15,
ymax = 0,
ymin = -25,
grid=major,
legend style={
at={(-0.25,1.6)},
anchor=north west}
]

\addplot[line width=1pt, color=black, style=solid] table[x index =0, y index =2] {fig/MIMO_channel_pattern_DDM.dat};
\label{p1}




\end{axis}

\end{tikzpicture}
\caption{Normalized power of the first subcarrier of an exemplary frequency selective ECFR plotted over the OFDM symbol index $\mu$.  \label{fig:MIMO_channel_pattern} }
\end{center}
\end{figure}
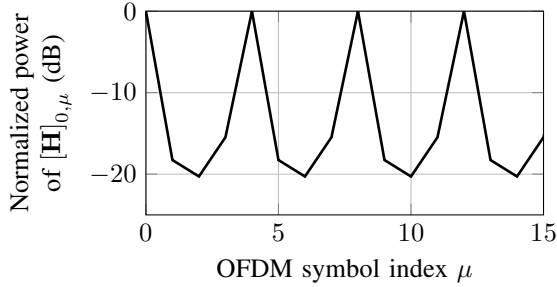

The same analysis for the employed frequency selective channel model \cite{Lang_RDM_JP} is presented in the following. For this model, the magnitude values of $\left[\m{H} \right]_{0,\mu}$ for an exemplary \ac{ecfr} are shown in Fig.~\ref{fig:MIMO_channel_pattern}.  Since the real-valued magnitude rather than the complex-valued amplitude values are shown, a period of 4 \ac{ofdm} symbols is observed. Within these 4 \ac{ofdm} symbols one can observe a pattern where 3 \ac{ofdm} symbols experience a rather strong attenuation, making a successful transmission difficult. Consequently, robustifying the communication is necessary.

\subsubsection{Increasing Robustness of the Communication}
\label{sec:Robust_DDM}

In wireless communications, bad channel conditions are usually tackled by adding redundancy to the transmit data. As discussed in \cite{Lang_RDM_JP}, redundancy may be added using an adequate channel code. Here, we face the same problem as in \cite{Lang_RDM_JP} for \ac{rdm}, where a usually sufficing convolutional code with code rate $r=1/2$ was not powerful enough to overcome the interference pattern in the \acp{ecfr}. 


The approach proposed in this work is similar to the one used in \cite{Lang_RDM_JP}. We propose adding redundancy to the transmit data by sending the same subcarrier symbols on $4$ consecutive \ac{ofdm} symbols in combination with a channel code with a code rate of $r=1/2$. The former one can be implemented with very low complexity and is inspired by space-time codes \cite{lu2000space, alamouti1998simple, tarokh1998space}, which utilize diversity gain to robustify  the transmission in difficult environments. The proposed approach also utilizes diversity gain, which will result in an increased \ac{ber} performance as will be shown later in this work. 

As discussed in \cite{Lang_RDM_JP}, adding redundancy  reduces the data rate, which may be seen as a moderate disadvantage for automotive \ac{ofdm} joint radar and communication systems, since the large bandwidths employed allow for very high data rates that exceed those of competitive radar waveforms significantly \cite{8897604}.

Transmitting the same subcarrier symbols on $4$ consecutive  \ac{ofdm} symbols is mathematically described by constructing  $\m{S}$ according to
\begin{align}
\m{S} =  \m{X} \m{B},  \label{equ:Comm_OFDM_00111}
\end{align}
with $\m{B}$ given as
\begin{align}
\m{B} &={} \left[ \begin{smallmatrix}
&  &  &  \\
1 & 1 & 1 & 1   & 0 & 0 & 0 & 0   & 0 & 0 & 0 & 0   & 0 & 0 & 0 & 0   &   \\
0 & 0 & 0 & 0   & 1 & 1 & 1 & 1   & 0 & 0 & 0 & 0   & 0 & 0 & 0 & 0   & \hdots \\
0 & 0 & 0 & 0   & 0 & 0 & 0 & 0   & 1 & 1 & 1 & 1   & 0 & 0 & 0 & 0   &  \\
0 & 0 & 0 & 0   & 0 & 0 & 0 & 0   & 0 & 0 & 0 & 0   & 1 & 1 & 1 & 1   &  \\
  & & \vdots & &   & &  \vdots & &   & & \vdots  &  &  & & \vdots  & & \ddots \\
\end{smallmatrix} \right]  \in \mathbb{R}^{(N_\text{sym}/4) \times N_\text{sym}}. \label{equ:IEEE_DDM_Bperm} 
\end{align} 

\subsection{Signal Model} 

For the mathematical description, $4$ consecutive \ac{ofdm} symbols within $\m{S}$ carrying the same subcarrier symbols are referred to as a bundle. Hence, $\m{S}$ contains $N_\text{sym}/4$ bundles indexed with $0 \leq \kappa < N_\text{sym}/4$. Within every bundle, the 4 \ac{ofdm} symbols are indexed with $0 \leq \gamma < 4$. This notation is visualized in Fig.~\ref{fig:Comm_setup_DDM}, which sketches the rows and columns of $\m{S}$. The columns of $\m{S}$ are denoted as $\ve{s}_{\kappa, \gamma} \in \mathbb{C}^{N_\text{c}}$ with $\kappa$ and $\gamma$ indicating the corresponding bundle and the \ac{ofdm} symbol within this bundle, respectively. The columns of the matrix $\m{X}$ are denoted as $\ve{x}_{\kappa} \in \mathbb{C}^{N_\text{c}}$.  This way, it holds that $\ve{x}_{\kappa} = \ve{s}_{\kappa, \gamma} $ for $0 \leq \gamma < 4$. 
 
Next, a mathematical model of the received time domain \ac{ofdm} symbols is derived. This model is similar to \eqref{equ:MIMO_OFDM_011} except that it applies to the communication receiver rather than the radar receiver. Hence, it uses the \ac{ecfr} discussed for the communication task. With $\m{P}_k = \text{diag}\left( \ve{p}_k \right)$, \eqref{equ:MIMO_OFDM_010}, \eqref{equ:Comm_OFDM_ECIR_002}, \eqref{equ:Comm_OFDM_ECIR_002qw}, and  \eqref{equ:Comm_OFDM_00111}, this model is given by
\begin{align}
	\m{Y}_{\text{tf,ts}} &={} \sum_{k=0}^{N_\text{Tx}-1} \m{F}_{N_\text{c}}^{-1} \m{P}_k \m{S}_k  \m{\Lambda} + \m{N} \label{equ:ECFR1_OFDM_011} \\
	&={} \sum_{k=0}^{N_\text{Tx}-1} \m{F}_{N_\text{c}}^{-1} \m{P}_k \m{S}  \m{D}_{N_\text{sym}} \left( \frac{\Delta \psi_k}{2 \pi} \right) \m{\Lambda}  + \m{N} \\
	 &={}  \m{F}_{N_\text{c}}^{-1} \left( \m{H} \odot \m{S} \right) \m{\Lambda} + \m{N}\\
	 &={}  \m{F}_{N_\text{c}}^{-1} \left( \m{H} \odot \left( \m{X} \m{B} \right) \right) \m{\Lambda} + \m{N}.\label{equ:ECFR2_OFDM_011}
\end{align}
Similar to \cite{Lang_RDM_JP}, the diagonal matrix $\m{\Lambda} \in \mathbb{C}^{N_\text{sym} \times N_\text{sym}}$ models the \acp{cpe}. Its diagonal elements are given by $\text{e}^{\text{j} \varphi_{\kappa, \gamma}}$ with $ \varphi_{\kappa, \gamma}$ representing the unknown \ac{cpe} for the $\gamma$th \ac{ofdm} symbol within the $\kappa$th bundle. The matrix $\m{N} \in \mathbb{C}^{N_\text{c} \times N_\text{sym}}$ represents zero-mean white Gaussian measurement noise, whose uncorrelated elements have a variance of $\sigma_\text{n}^2$.

Recall that the \ac{ecfr} $\m{H}$ shows a period of 8 columns. Within this period, the columns of $\m{H}$ show a pattern where the first 4 columns are equal with the following 4 columns when inverting all signs. This pattern is accounted for by inverting the signs of the received signals for the corresponding \ac{ofdm} symbols as described in the following. 
Let $\ve{y}_{\kappa, \gamma} \in \mathbb{C}^{N_\text{c}}$ be the column of $\m{Y}_{\text{tf,ts}}$ corresponding to the $\gamma$th \ac{ofdm} symbol within the $\kappa$th bundle, and let $\ve{z}_{\kappa, \gamma} \in \mathbb{C}^{N_\text{c}}$ be its \ac{dft} transform up to a sign inversion in case of odd vales of $\kappa$
\begin{align}
	\ve{z}_{\kappa, \gamma} &={} \left(-1\right)^{\kappa} \m{F}_{N_\text{c}} \ve{y}_{\kappa, \gamma}. \label{equ:IEEE_CIR032kappa} 
\end{align}
The follow-up receiver signal processing is based on $\ve{z}_{\kappa, \gamma}$ rather than on $\ve{y}_{\kappa, \gamma}$, which allows describing the \ac{ecfr} $\m{H}$ by only 4 columns. These 4 columns will be denoted as $\ve{h}_{\gamma} \in \mathbb{C}^{N_\text{c}}$ for $\gamma = 0, \hdots, 3$. From now on, these 4 vectors are referred to \ac{ecfr} for the sake of simplicity. A diagonal matrix with the \ac{ecfr} is defined as $\m{H}_{\gamma} = \text{diag}\left( \ve{h}_{\gamma}  \right)\in \mathbb{C}^{N_\text{c} \times N_\text{c}}$.

The \ac{ecfr} $\ve{h}_{\gamma}$ transformed into the time domain is denoted as \ac{ecir} $\ve{g}_{\gamma}\in \mathbb{C}^{N_\ve{g}}$, whose length $N_\ve{g}$ corresponds to the length of the \ac{cir} $\ve{f}_{k}$ of  $N_\ve{f} = N_\ve{g} = 256$. The \ac{ecir} and the \ac{ecfr} are connected via
\begin{align}
\ve{h}_{\gamma} = \m{F}_{N_\text{c}} \m{B}_{\text{zp}} \ve{g}_{\gamma}. \label{equ:DDM_001}
\end{align}
The introduced definitions and notations allow simplifying the model in \eqref{equ:ECFR1_OFDM_011}--\eqref{equ:ECFR2_OFDM_011} as
\begin{align}
	\ve{z}_{\kappa, \gamma} &={} \m{H}_{\gamma} \ve{s}_{\kappa, \gamma} \text{e}^{\text{j} \varphi_{\kappa, \gamma}} + \ve{n}_{\kappa, \gamma} \label{equ:MODEL_DDM_prea001} \\
	&={} \m{H}_{\gamma}  \ve{x}_{\kappa} \text{e}^{\text{j} \varphi_{\kappa, \gamma}} + \ve{n}_{\kappa, \gamma} , \label{equ:MODEL_DDM_prea002} 
\end{align}
where $\ve{n}_{\kappa,\gamma}\in \mathbb{C}^{N_\text{c}}$ is a white Gaussian noise vector given by the \ac{dft} of the corresponding columns of $\m{N}$. The alternating sign considered in \eqref{equ:IEEE_CIR032kappa} is ignored for the noise, since it does not affect its statistics. The covariance matrix of $\ve{n}_{\kappa,\gamma}$ is given by $\m{C}_{\ve{n}\ve{n}}=N_\text{c} \sigma_\text{n}^2 \m{I}^{N_\text{c}}$. 

The matrix $\m{S}$ contains $N_\text{pr}$ preamble \ac{ofdm} symbols in the first $N_\text{pr}$ columns, which are used for channel estimation later in this work. Thus, $\m{X}$ contains $N_\text{pr}/4$ preamble \ac{ofdm} symbols in the first $N_\text{pr}/4$ columns, where it is assumed that $N_\text{pr}$ is a multiple of 4. Let $\ve{x}_\text{pr}\in \mathbb{C}^{N_\text{c}}$ be the preamble \ac{ofdm} symbol in frequency domain such that $\ve{x}_{\kappa} = \ve{x}_\text{pr}$ for $0 \leq \kappa < N_\text{pr}/4$. 

For $\kappa \geq N_\text{pr}/4$, $\ve{x}_{\kappa}$ consists of $N_\text{p}$ pilot and $N_\text{d}$ data subcarriers such that $N_\text{c} = N_\text{p} + N_\text{d}$. The symbols transmitted on the pilot subcarriers are known to the receiver and are used for synchronization. The variance of the assumed uncorrelated data subcarriers is denoted as $\sigma_\text{d}^2$, which is usually normalized such that $\sigma_\text{d}^2=1$.

Dividing subcarriers into data subcarriers and pilot subcarriers not only applies to $\ve{x}_{\kappa}$ but also to many other variables such as $\ve{s}_{\kappa, \gamma}$, $\ve{z}_{\kappa, \gamma}$, $\ve{n}_{\kappa, \gamma}$, $\ve{h}_{\gamma}$, and $\m{H}_{\gamma}$. In the following, the superscript 'p' indicates the pilot subcarriers only, and the superscript 'd' refers the data subcarriers only. For instance, the sub-vector of $\ve{z}_{\kappa, \gamma}$ containing the pilot subcarriers only is denoted as $\ve{z}_{\kappa, \gamma}^\text{p}\in \mathbb{C}^{N_\text{p}}$, and the sub-vector containing the data subcarriers only is given by $\ve{z}_{\kappa, \gamma}^\text{d}\in \mathbb{C}^{N_\text{d}}$.

\subsection{Channel Estimation} \label{sec:channel_estimation}

The \ac{ofdm} preamble symbols are utilized for channel estimation. As a preparatory step, these symbols are synchronized to account for a potential \ac{cpe}, e.g., caused by a relative velocity between transmitter and receiver. After that, the \ac{ecir} is estimated, which is then transformed into an estimate of the \ac{ecfr}.

\newcommand\DELTA{0.5}
\begin{figure}[!t]
\centering
\begin{tikzpicture}[scale=0.8]
    \draw [<->,thick] (0,4.2) node (yaxis) [left] {}
        |- (6.2,0) node (xaxis) [below] {$\mu$};
    \node[left, anchor=south, rotate=90] at (0,2.0) {Subcarriers};
    \node[below] at (3,0) {OFDM Symbols};
	\draw[step=\DELTA,black,thin,xshift=0cm,yshift=0cm] (0,0) grid (6,4);
	\draw[draw=none, fill=gray, opacity=0.2] (0,0) rectangle ++(\DELTA, 4);
	\draw[draw=none, fill=gray, opacity=0.4] (\DELTA,0) rectangle ++(\DELTA, 4);
	\draw[draw=none, fill=gray, opacity=0.2] (2*\DELTA,0) rectangle ++(\DELTA, 4);
	\draw[draw=none, fill=gray, opacity=0.4] (3*\DELTA,0) rectangle ++(\DELTA, 4);


  
\draw[draw=none, fill=blue, opacity=0.2] (2,1) rectangle ++(4,\DELTA);
\draw[draw=none, fill=blue, opacity=0.2] (2,2.5) rectangle ++(4,\DELTA);
\draw [->,blue, thick] (6.1, 2.75) to [out=0,in=130] (6.5, 2.25);
\draw [->,blue, thick] (6.1, 1.25) to [out=0,in=-130] (6.5, 1.75);
\node[right, anchor=center, rotate=90, color = blue] at (6.8, 2.0) {Pilot subcarriers};



\draw[decorate,decoration={brace,amplitude=5pt}, thick]  (0,5.6) -- (2,5.6);
\node (pr0) at (1, 6.1) {\footnotesize $\kappa = 0$};
\draw[decorate,decoration={brace,amplitude=5pt}, thick]  (2,5.6) -- (4,5.6);
\node (s0) at (3, 6.1) {\footnotesize $\kappa = 1$};
\draw[decorate,decoration={brace,amplitude=5pt}, thick]  (4,5.6) -- (6,5.6);
\node (s1) at (5, 6.1) {\footnotesize $\kappa = 2$};

\node[align=center] (cu0) at (3, 6.5) {\footnotesize Bundles};

\node[right, rotate=90] (g5) at (0.25, 4.4) {\footnotesize $\gamma = 0$};
\node[right, rotate=90] (g6) at (0.75, 4.4) {\footnotesize $\gamma = 1$};
\node[right, rotate=90] (g7) at (1.25, 4.4) {\footnotesize $\gamma = 2$};
\node[right,  rotate=90] (g8) at (1.75, 4.4) {\footnotesize $\gamma = 3$};

\node[right, rotate=90] (g1) at (2.25, 4.4) {\footnotesize $\gamma = 0$};
\node[right, rotate=90] (g2) at (2.75, 4.4) {\footnotesize $\gamma = 1$};
\node[right, rotate=90] (g3) at (3.25, 4.4) {\footnotesize $\gamma = 2$};
\node[right, rotate=90] (g4) at (3.75, 4.4) {\footnotesize $\gamma = 3$};

\draw[->, to path={-| (\tikztotarget)}]
  (2.25,4.2) -- (g1);
  \draw[->, to path={-| (\tikztotarget)}]
  (2.75,4.2) -- (g2);
  \draw[->, to path={-| (\tikztotarget)}]
  (3.25,4.2) -- (g3);
  \draw[->, to path={-| (\tikztotarget)}]
  (3.75,4.2) -- (g4);
  \draw[->, to path={-| (\tikztotarget)}]
  (0.25,4.2) -- (g5);
  \draw[->, to path={-| (\tikztotarget)}]
  (0.75,4.2) -- (g6);
  \draw[->, to path={-| (\tikztotarget)}]
  (1.25,4.2) -- (g7);
  \draw[->, to path={-| (\tikztotarget)}]
  (1.75,4.2) -- (g8);
 
 \node[right, anchor=center, align=center] at (5.0, 5.0) {$\hdots$};
 
 \draw[decorate,decoration={brace,amplitude=5pt}, thick, anchor=center, rotate=180]  (-2,0.6) -- (0,0.6);
 \node[align=center] (cu0) at (1, -1.1) {\footnotesize Preamble symbols};
  

\end{tikzpicture}
\caption{Exemplary allocation of preamble OFDM symbols and pilot subcarriers in $\m{S}$ for DDM.}
\label{fig:Comm_setup_DDM}
\end{figure}
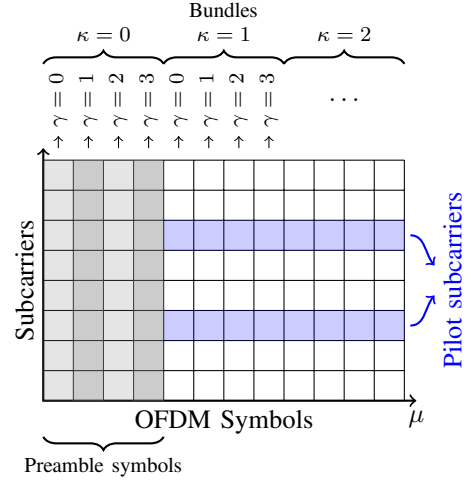

\subsubsection{Synchronization of the Preamble OFDM Symbols} 

The model in \eqref{equ:MODEL_DDM_prea002}  serves as a basis for synchronization, however, as usual in channel estimation, the roles of the channel and the preamble \ac{ofdm} symbols are reversed. This yields
\begin{align}
	\ve{z}_{\kappa, \gamma} &={}  \m{X}_\text{pr}  \ve{h}_{\gamma} \text{e}^{\text{j} \varphi_{\kappa, \gamma}} + \ve{n}_{\kappa, \gamma}, \label{equ:IEEE_DDM_prea001} 
\end{align}
where $\m{X}_\text{pr} = \text{diag}\left( \ve{x}_\text{pr}  \right)\in \mathbb{C}^{N_\text{c} \times N_\text{c}}$.  Without loss of generality, $\varphi_{0,\gamma}$ is set to $0$ and we estimate the \ac{cpe} within $\ve{z}_{\kappa, \gamma}$ with respect to $\ve{z}_{0, \gamma}$ via \cite{classen1994frequency, Diss_Hofbauer, huemer2002simulation }
\begin{align}
\widehat{\varphi}_{\kappa,\gamma} = \text{arg} \left( \ve{z}_{0, \gamma}^H  \ve{z}_{\kappa,\gamma} \right),	 \label{equ:IEEE__DDM_prea_pilot006a} 
\end{align}
whereas $\veh{\cdot}$ indicates that $\widehat{\varphi}_{\kappa,\gamma} $ is an estimate of $\varphi_{\kappa, \gamma} $.  This estimation procedure is repeated for $0 \leq \gamma < 4$ and for $0 \leq \kappa < N_\text{pr}/4$. The estimated \acp{cpe} are then used to synchronize $\ve{z}_{\kappa, \gamma}$ according to
\begin{align}
\vef{z}_{\kappa,\gamma} &={} \ve{z}_{\kappa,\gamma} \cdot \text{e}^{-\text{j} \widehat{\varphi}_{\kappa,\gamma} }.	 \label{equ:IEEE__DDM_prea_pilot007} 
\end{align}
After synchronization, averaging within one bundle yields
\begin{align}
	\bar{\vef{z}}_{\gamma} = \frac{1}{N_\text{pr}/4} \sum_{\kappa=0}^{N_\text{pr}/4-1}  \vef{z}_{\kappa,\gamma}.  \label{equ:IEEE_CIR015ddd}	
\end{align}
With \eqref{equ:DDM_001} and \eqref{equ:IEEE_CIR015ddd}, \eqref{equ:IEEE_DDM_prea001} can be approximated by 
\begin{align}
	\bar{\vef{z}}_{\gamma} &\approx{}  \underbrace{\m{X}_\text{pr} \m{F}_{N_\text{c}} \m{B}_{\text{zp}}}_{\m{M}_\text{pr}} \ve{g}_{\gamma} + \underbrace{\frac{1}{N_\text{pr}/4} \sum_{\kappa=0}^{N_\text{pr}/4-1}  \ve{n}_{\kappa,\gamma}}_{\ve{n}_{ \gamma}} \\
	&={} \m{M}_\text{pr} \ve{g}_{\gamma} + \ve{n}_{ \gamma}, \label{equ:IEEE_CIR0329}
\end{align}
which is the basis for the following estimation procedure.

\subsubsection{ECIR/ECFR Estimation} 

Employing the commonly used \ac{blue} \cite{Kay-Est.Theory, salehi2007digital, Lang_Asilomar_2014, Diss_Lang_Oliver} on \eqref{equ:IEEE_CIR0329}, an estimate of the \ac{ecir} $\ve{g}_{\gamma}$ follows as
\begin{align}
	\veh{g}_{\gamma} =  \left(\m{M}_\text{pr}^H \m{M}_\text{pr}\right)^{-1} \m{M}_\text{pr}^H \bar{\vef{z}}_{\gamma}.  \label{equ:IEEE_CIR015}	
\end{align}
The estimated \acp{ecir} in \eqref{equ:IEEE_CIR015} can be transformed into an estimate of the corresponding \acp{ecfr} $\ve{h}_{\gamma}$ via 
\begin{align}
\veh{h}_{\gamma} = \m{F}_{N_\text{c}} \m{B}_{\text{zp}} \veh{g}_{\gamma}. \label{equ:DDM_002}
\end{align}
The matrix representation of the estimate $\veh{h}_{\gamma}$ is defined as $\mh{H}_{\gamma} = \text{diag}\left( \veh{h}_{\gamma}  \right)\in \mathbb{C}^{N_\text{c} \times N_\text{c}}$. This procedure is repeated for all indexes $\gamma = 0, \hdots, 3$.

\subsection{Synchronization of OFDM Symbols for $\kappa \geq N_\text{pr}/4$} 

Considering only the pilot subcarriers of the model in \eqref{equ:MODEL_DDM_prea001} and replacing the \ac{ecfr} by its estimate yields
\begin{align}
	\ve{z}_{\kappa, \gamma}^\text{p} &\approx{} \mh{H}_{\gamma}^\text{p}  \underbrace{\ve{s}_{\kappa, \gamma}^\text{p} \text{e}^{\text{j} \varphi_{\kappa, \gamma}}}_{\ve{t}_{\kappa, \gamma}^\text{p}} + \ve{n}_{\kappa, \gamma}^\text{p} \\
	&={} \mh{H}_{\gamma}^\text{p} \ve{t}_{\kappa, \gamma}^\text{p} + \ve{n}_{\kappa, \gamma}^\text{p}, \label{equ:IEEE_pilot000} 
\end{align}
where $\ve{t}_{\kappa, \gamma}^\text{p} = \ve{s}_{\kappa, \gamma}^\text{p} \text{e}^{\text{j} \varphi_{\kappa, \gamma}} \in \mathbb{C}^{N_\text{p}}$ represents \ac{cpe} distorted pilot symbols  \cite{Hofbauer20_2, Hofbauer20_2a}. Employing the commonly used \ac{lmmse} estimator \cite{Kay-Est.Theory, Diss_Hofbauer,  huemer2011classical, huemer2002simulation, classen1994frequency, Diss_Lang_Oliver, Lang_RDM_JP} on \eqref{equ:IEEE_pilot000} yields
\begin{align}
\veh{t}_{\kappa, \gamma}^\text{p} &={}  \left( \left(\mh{H}_{\gamma}^\text{p} \right)^H \mh{H}_{\gamma}^\text{p} + N_\text{c} \sigma_\text{n}^2 \m{C}_{\ve{t}\ve{t}}^{-1} \right)^{-1}  \left(\mh{H}_{\gamma}^\text{p} \right)^H \ve{z}_{\kappa, \gamma}^\text{p}.	 \label{equ:IEEE_pilot005a} 
\end{align}
There, $\m{C}_{\ve{t}\ve{t}}\in \mathbb{C}^{N_\text{p} \times N_\text{p}}$ denotes the covariance matrix of $\ve{t}_{\kappa, \gamma}^\text{p}$ and it is assumed to be a diagonal matrix for simplicity. The diagonal elements of $\m{C}_{\ve{t}\ve{t}}$ represent the pilot symbols' average power (averaged over the \ac{ofdm} symbols). The estimate \ac{cpe} distorted pilot symbols $\veh{t}_{\kappa, \gamma}^\text{p}$ in \eqref{equ:IEEE_pilot005a} 
feature the error covariance matrix  \cite{Diss_Hofbauer, Kay-Est.Theory, Diss_Lang_Oliver}
\begin{align}
\m{C}_{ee}^\text{p} &={} N_\text{c} \sigma_\text{n}^2 \left( \left(\mh{H}_{\gamma}^\text{p} \right)^H \mh{H}_{\gamma}^\text{p} + N_\text{c} \sigma_\text{n}^2 \m{C}_{\ve{t}\ve{t}}^{-1} \right)^{-1} .	 \label{equ:IEEE_pilot006} 
\end{align}
Comparing the estimates $\veh{t}_{\kappa, \gamma}^\text{p}$ in \eqref{equ:IEEE_pilot005a} with the known transmitted pilot symbols $\ve{s}_{\kappa, \gamma}^\text{p}$ allows estimating the \ac{cpe} for every $\kappa \geq N_\text{pr}/4$ and for $ 0 \leq \gamma < 4$ according to \cite{classen1994frequency, Diss_Hofbauer, huemer2002simulation}
\begin{align}
\widehat{\varphi}_{\kappa, \gamma} = \text{arg} \left( \left( \ve{s}_{\kappa, \gamma}^\text{p} \right)^H\m{W}\, \veh{t}_{\kappa, \gamma}^\text{p} \right).	 \label{equ:IEEE_pilot006a} 
\end{align}
Here, the diagonal matrix $\m{W}\in \mathbb{C}^{N_\text{p} \times N_\text{p}}$ weights the estimated pilot subcarriers based on their estimation accuracy indicated by $\m{C}_{ee}^\text{p}$ in \eqref{equ:IEEE_pilot006} \cite{Diss_Hofbauer, Lang_RDM_JP}. 
Finally, the estimated \acp{cpe} are used for de-rotating the received data subcarriers according to
\begin{align}
\vef{z}_{\kappa, \gamma}^\text{d} &={}  \ve{z}_{\kappa, \gamma}^\text{d} \cdot \text{e}^{-\text{j} \widehat{\varphi}_{\kappa, \gamma} }.	 \label{equ:IEEE_pilot007} 
\end{align}

\subsection{Data Estimation} 

Recall that the same data symbols are transmitted over 4 consecutive \ac{ofdm} symbols according to \eqref{equ:Comm_OFDM_00111}. Hence, the 4 vectors $\vef{z}_{\kappa, 0}^\text{d}, \cdots, \vef{z}_{\kappa, 3}^\text{d}$ are used to estimate the data symbols in $\ve{x}_{\kappa}^\text{d} $. The connection between these vectors is given by
\begin{align}
	\underbrace{\begin{bmatrix}
	\vef{z}_{\kappa, 0}^\text{d} \\
	\vef{z}_{\kappa, 1}^\text{d} \\
	\vef{z}_{\kappa, 2}^\text{d} \\
	\vef{z}_{\kappa, 3}^\text{d}  
	\end{bmatrix}}_{\vef{z}_{\kappa}^\text{d}\in \mathbb{C}^{4N_\text{d}}}
	&={}  \underbrace{\begin{bmatrix}  
\mh{H}_{0}^\text{d}  \\
\mh{H}_{1}^\text{d}  \\ 
\mh{H}_{2}^\text{d}  \\  
\mh{H}_{3}^\text{d}  \end{bmatrix}}_{\mh{H}^\text{d}\in \mathbb{C}^{4N_\text{d}\times N_\text{d}}} 
\ve{x}_{\kappa}^\text{d} + \underbrace{\begin{bmatrix}
	\ve{n}_{\kappa,0}^\text{d} \\
	\ve{n}_{\kappa,1}^\text{d} \\
	\ve{n}_{\kappa,2}^\text{d} \\
	\ve{n}_{\kappa,3}^\text{d} 
	\end{bmatrix}}_{\ve{n}_{\kappa}^\text{d}\in \mathbb{C}^{4N_\text{d}}}   \label{equ:IEEE_CIR032d}  \\ \nonumber \\
\vef{z}_{\kappa}^\text{d} &={} \mh{H}^\text{d} \ve{x}_{\kappa}^\text{d} + \ve{n}_{\kappa}^\text{d}.	 \label{equ:IEEE_CIR032c} 
\end{align}
The \ac{lmmse} estimator for $\ve{x}_{\kappa}^\text{d}$ can be derived as \cite{Diss_Hofbauer, Kay-Est.Theory,  huemer2011classical} 
\begin{align}
\widehat{\ve{x}}_{\kappa}^\text{d} &={}  \left( \left(\mh{H}^\text{d} \right)^H \mh{H}^\text{d} + \frac{N_\text{c} \sigma_\text{n}^2}{\sigma_\text{d}^2} \m{I}^{N_\text{d}} \right)^{-1}  \left(\mh{H}^\text{d} \right)^H \vef{z}_{\kappa}^\text{d},	 \label{equ:IEEE_CIR032e} 
\end{align}
 representing the final estimate of the data symbols.


\section{BER Performance Comparison} \label{sec:BER_performance}


In this section, the \ac{ber} performance of the proposed \ac{mimo} \ac{ofdm} system using \ac{ddm} is compared with that of a \ac{siso} \ac{ofdm} system and with that of \ac{mimo} \ac{ofdm} systems utilizing \ac{esi}, \ac{rdm}, and \ac{dsi} investigated in \cite{Hakobyan_A_novel_OFDM_MIMO_wo_CS}. For the latter one, it is assumed that the receiver knows the assignment of the subcarrier sets to the individual Tx antennas.
All considered systems employ the system parameters listed in Tab.~\ref{Tab:com_sys_paramters} except for the \ac{siso} \ac{ofdm} system for which $N_\text{Tx} = 1$. 

Additional processing blocks, e.g., the channel coder/decoder \cite{Viterbi, salehi2007digital, Diss_Hofbauer}, the mapper/demapper \cite{allpress2004exact, Haselmayr_LLRs, Lang_Asilomar_2016_LLRs}, interleaver/deinterleaver, randomized \ac{cir} generation \cite{Rappaport_SISO_Channel, Rappaport_SISO_Channel_JP, Code_MIMO_Channels}, correspond to those utilized in \cite{Lang_RDM_JP} such that a detailed description can be omitted in this work.

The simulations are carried out for fixed values of $E_\text{b}/N_\text{0}$, where $E_\text{b}$ is the average energy per bit of information, and where $N_\text{0}/2$ is the double-sided noise power spectral density of a bandpass noise signal \cite{Diss_Hofbauer}. To obtain the desired value of $E_\text{b}/N_\text{0}$ the noise variance $\sigma_\text{n}^2$ of the complex-valued \ac{awgn} at the receiver input is chosen according to \cite{Miller_compl_BB, Diss_Hofbauer}
\begin{align}
	\sigma_\text{n}^2 = \frac{P_\text{s}}{(E_\text{b}/N_\text{0})br  \zeta \nu }. \label{equ:IEEE_CIR015a}	
\end{align}
There, $P_\text{s}$ represents the average signal power per time-domain sample measured at the receiver input. Moreover, $b$ is the number of bits per data symbol ($b=2$ for \ac{qpsk}), $r$ represents the code rate of the channel code, and  $\zeta= N_\text{c} / (N_\text{cp} + N_\text{c})$ accounts for the time domain samples in the \ac{cp}. The parameter $\nu$ accounts for the additional redundancy discussed in Sec.~\ref{sec:Robust_DDM}. Thus, $\nu=\frac{1}{4}$ for the \ac{mimo} \ac{ofdm} system employing \ac{ddm}. \ac{rdm} adds a  similar redundancy \cite{Lang_RDM_JP}, such that $\nu=\frac{1}{4}$ is chosen also for the \ac{mimo} \ac{ofdm} system utilizing \ac{rdm}. All other considered systems do not add any additional redundancy such that $\nu=1$ is chosen for them. As a result, \ac{mimo} \ac{ofdm} systems employing \ac{ddm} and \ac{rdm} observe a higher noise variance $\sigma_\text{n}^2$. 

The \ac{ber} curves are simulated for three simulation scenarios detailed in the following.

\subsubsection{Perfect Channel Knowledge; Perfect Synchronization}

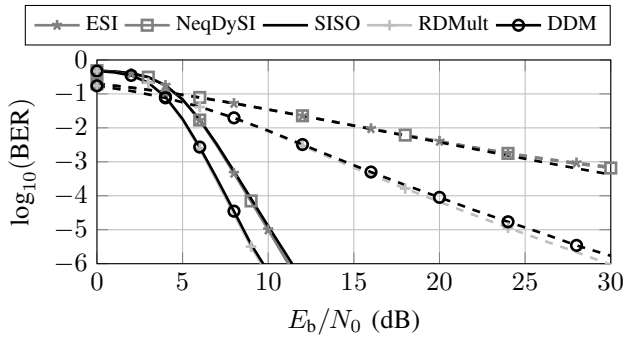
\begin{figure}[!t]
\begin{center}
\begin{tikzpicture}
\begin{axis}[compat=newest, 
width=.98\columnwidth, height = .5\columnwidth, grid, xlabel={$E_\text{b}/N_0$ (dB)}, 
ylabel={$\text{log}_{10}(\text{BER})$}, 
legend pos=south east, 
legend cell align=left,
legend columns=5, 
xmin = 0,
xmax = 30,
ymin = -6,
ymax = 0,
ytick={0, -1, -2, -3, -4, -5, -6},
legend style={
at={(1.0, 1.27)},
anchor=north east,font=\tiny}
]

\addplot[line width=1pt][color=gray, solid, every mark/.append style={solid},mark=star,mark repeat = 2] table[x index =0, y index =1] {SimData/FP_ESI_Ntx4_Nc1024_Nsym256_R12_CI1_CK0_CD2_Nbps2_Nprea4_Npi16.dat};
\addlegendentry{{\footnotesize ESI}}

\addplot[line width=1pt][color=gray, solid, every mark/.append style={solid},mark=square,mark repeat = 3] table[x index =0, y index =1] {SimData/FP_DSI_Ntx4_Nc1024_Nsym256_R12_CI1_CK0_CD2_Nbps2_Nprea4_Npi16.dat};
\addlegendentry{{\footnotesize NeqDySI}}

\addplot[line width=1pt][color=black, solid] table[x index =0, y index =1] {SimData/FP_SISO_Ntx4_Nc1024_Nsym256_R12_CI1_CK0_CD2_Nbps2_Nprea4_Npi16.dat};
\addlegendentry{{\footnotesize SISO}}

\addplot[line width=1pt][color=lightgray, solid, every mark/.append style={solid},mark=+,mark repeat = 3] table[x index =0, y index =1] {SimData/FP_RDMult_Ntx4_Nc1024_Nsym256_R12_CI1_CK0_CD2_Nbps2_Nprea4_Npi16.dat};
\addlegendentry{{\footnotesize RDMult}}

\addplot[line width=1pt][color=black, solid, every mark/.append style={solid},mark=o,mark repeat = 2] table[x index =0, y index =1] {SimData/FP2_DDM_Ntx4_Nc1024_Nsym256_R12_CI1_CK0_CD2_Nbps2_Nprea4_Npi16.dat};
\addlegendentry{{\footnotesize DDM}}

\addplot[line width=1pt][color=gray, dashed, every mark/.append style={solid},mark=star,mark repeat = 2] table[x index =0, y index =1] {SimData/FP_ESI_Ntx4_Nc1024_Nsym64_R1_CI1_CK0_CD2_Nbps2_Nprea4_Npi16.dat};

\addplot[line width=1pt][color=gray, dashed, every mark/.append style={solid},mark=square,mark repeat = 3] table[x index =0, y index =1] {SimData/FP_DSI_Ntx4_Nc1024_Nsym64_R1_CI1_CK0_CD2_Nbps2_Nprea4_Npi16.dat};

\addplot[line width=1pt][color=black, dashed] table[x index =0, y index =1] {SimData/FP_SISO_Ntx4_Nc1024_Nsym64_R1_CI1_CK0_CD2_Nbps2_Nprea4_Npi16.dat};

\addplot[line width=1pt][color=lightgray, dashed, every mark/.append style={solid},mark=+,mark repeat = 3] table[x index =0, y index =1] {SimData/FP_RDMult_Ntx4_Nc1024_Nsym64_R1_CI1_CK0_CD2_Nbps2_Nprea4_Npi16.dat};

\addplot[line width=1pt][color=black, dashed, every mark/.append style={solid},mark=o,mark repeat = 2] table[x index =0, y index =1] {SimData/FP_DDM_Ntx4_Nc1024_Nsym64_R1_CI1_CK0_CD2_Nbps2_Nprea4_Npi16.dat};

\end{axis}
\end{tikzpicture}
\caption{BER curves for the case of perfect synchronization and perfect channel knowledge. The solid lines are for coded, and the dashed lines are for uncoded transmission. 
\label{fig:BER_optimal} }
\end{center}
\end{figure}

In this first simulation, the receiver perfectly knows the channel between transmitter and receiver. The observed \ac{ber} curves for uncoded and coded transmission are shown in Fig.~\ref{fig:BER_optimal}. While for the uncoded case \ac{rdm} has a small  advantage in \ac{ber} performance over \ac{ddm}, both systems feature approximately the same \ac{ber} performance in the coded case and outperform the remaining systems by approximately $1.6\, \text{dB}$. This gain in \ac{ber} performance is a result of the diversity gain elaborated on in Sec.~\ref{sec:Robust_DDM}.  

As argued in \cite{Lang_RDM_JP}, granting this diversity gain also to \ac{mimo} systems utilizing \ac{esi} and \ac{dsi} by means of adding additional redundancy would increase their \ac{ber} performances as well at the cost of a reduced data rate. 

The remaining simulations are shown for coded transmission only, since uncoded transmission is not relevant for real-world applications. 

\subsubsection{Perfect Synchronization; Imperfect Channel Estimation based on Preamble OFDM Symbols}

Now, the channel is not perfectly known but rather estimated using the procedure derived in Sec.~\ref{sec:channel_estimation}. The channel estimation procedure for the \ac{siso} \ac{ofdm} system is described in \cite{Diss_Hofbauer, Lang_Asilomar_2014}. The channels for the \ac{mimo} \ac{ofdm} systems utilizing \ac{esi} and \ac{dsi} are estimated with the \ac{blue} \cite{Kay-Est.Theory, Diss_Lang_Oliver}, whose derivations are omitted in this work.

For a fair comparison by means of having the similar distortions on the channel estimates, all three systems shall have the same effective \ac{snr} for the averaged preamble \ac{ofdm} symbols \cite{Lang_RDM_JP}. Hence, the increased noise variance $\sigma_\text{n}^2$ for the \ac{mimo} \ac{ofdm} systems with \ac{ddm} and \ac{rdm} is compensated by employing $N_\text{pr}=16$ preamble \ac{ofdm} symbols, while the other systems use $N_\text{pr}=4$. 

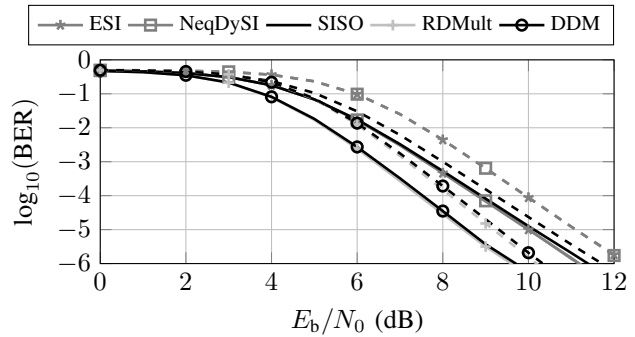
\begin{figure}[!t]
\begin{center}
\begin{tikzpicture}
\begin{axis}[compat=newest, 
width=.98\columnwidth, height = .5\columnwidth, grid, xlabel={$E_\text{b}/N_0$ (dB)}, 
ylabel={$\text{log}_{10}(\text{BER})$}, 
legend pos=south east, 
legend cell align=left,
legend columns=5, 
xmin = 0,
xmax = 12,
ymin = -6,
ymax = 0,
ytick={0, -1, -2, -3, -4, -5, -6},
legend style={
at={(1.0, 1.27)},
anchor=north east,font=\tiny}
]

\addplot[line width=1pt][color=gray, solid, every mark/.append style={solid},mark=star,mark repeat = 2] table[x index =0, y index =1] {SimData/FP_ESI_Ntx4_Nc1024_Nsym256_R12_CI1_CK0_CD2_Nbps2_Nprea4_Npi16.dat};
\addlegendentry{{\footnotesize ESI}}

\addplot[line width=1pt][color=gray, solid, every mark/.append style={solid},mark=square,mark repeat = 3] table[x index =0, y index =1] {SimData/FP_DSI_Ntx4_Nc1024_Nsym256_R12_CI1_CK0_CD2_Nbps2_Nprea4_Npi16.dat};
\addlegendentry{{\footnotesize NeqDySI}}

\addplot[line width=1pt][color=black, solid] table[x index =0, y index =1] {SimData/FP_SISO_Ntx4_Nc1024_Nsym256_R12_CI1_CK0_CD2_Nbps2_Nprea4_Npi16.dat};
\addlegendentry{{\footnotesize SISO}}

\addplot[line width=1pt][color=lightgray, solid, every mark/.append style={solid},mark=+,mark repeat = 3] table[x index =0, y index =1] {SimData/FP_RDMult_Ntx4_Nc1024_Nsym256_R12_CI1_CK0_CD2_Nbps2_Nprea4_Npi16.dat};
\addlegendentry{{\footnotesize RDMult}}

\addplot[line width=1pt][color=black, solid, every mark/.append style={solid},mark=o,mark repeat = 2] table[x index =0, y index =1] {SimData/FP2_DDM_Ntx4_Nc1024_Nsym256_R12_CI1_CK0_CD2_Nbps2_Nprea4_Npi16.dat};
\addlegendentry{{\footnotesize DDM}}

\addplot[line width=1pt][color=gray, dashed, every mark/.append style={solid},mark=star,mark repeat = 2] table[x index =0, y index =1] {SimData/FP_ESI_Ntx4_Nc1024_Nsym256_R12_CI1_CK1_CD2_Nbps2_Nprea4_Npi16.dat};

\addplot[line width=1pt][color=gray, dashed, every mark/.append style={solid},mark=square,mark repeat = 3] table[x index =0, y index =1] {SimData/FP_DSI_Ntx4_Nc1024_Nsym256_R12_CI1_CK1_CD2_Nbps2_Nprea4_Npi16.dat};

\addplot[line width=1pt][color=black, dashed] table[x index =0, y index =1] {SimData/FP_SISO_Ntx4_Nc1024_Nsym256_R12_CI1_CK1_CD2_Nbps2_Nprea4_Npi16.dat};

\addplot[line width=1pt][color=lightgray, dashed, every mark/.append style={solid},mark=+,mark repeat = 3] table[x index =0, y index =1] {SimData/FP2_RDMult_Ntx4_Nc1024_Nsym256_R12_CI1_CK1_CD2_Nbps2_Nprea16_Npi16.dat};

\addplot[line width=1pt][color=black, dashed, every mark/.append style={solid},mark=o,mark repeat = 2] table[x index =0, y index =1] {SimData/FP2_DDM_Ntx4_Nc1024_Nsym256_R12_CI1_CK1_CD2_Nbps2_Nprea16_Npi16.dat};

%

\end{axis}
\end{tikzpicture}
\caption{BER curves for the case of perfect synchronization, code rate $r=1/2$, and for perfect channel knowledge (solid) and for imperfect channel estimation (dashed). 
\label{fig:BER_channel_estimation} }
\end{center}
\end{figure}

The resulting \ac{ber} curves are shown in Fig.~\ref{fig:BER_channel_estimation}. This figure also visualizes the simulation results for the case of perfect channel knowledge from Fig.~\ref{fig:BER_optimal} as reference. While the \ac{mimo} systems are less prone to imperfect channel knowledge, the loss in performance for all three systems is moderate\footnote{We note that in practice, not only the ECIRs but also the underlying CIRs may become highly time-varying for the assumed relative velocity between $\pm 60\,\text{m/s}$. This may entail the necessity of more frequent channel estimation or advanced channel tracking algorithms, whose analysis is beyond the scope of this work.}. 

\subsubsection{Perfect Channel Knowledge; Imperfect Synchronization using Pilot Subcarriers}

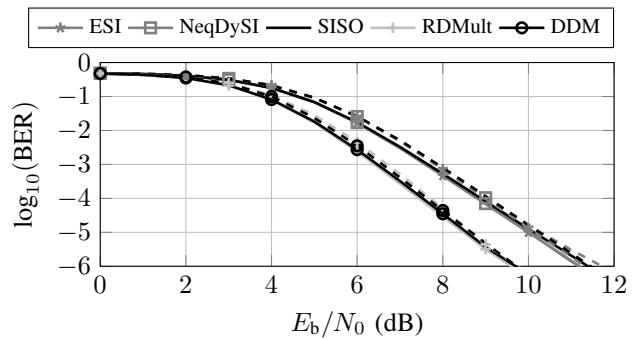
\begin{figure}[!t]
\begin{center}
\begin{tikzpicture}
\begin{axis}[compat=newest, 
width=.98\columnwidth, height = .5\columnwidth, grid, xlabel={$E_\text{b}/N_0$ (dB)}, 
ylabel={$\text{log}_{10}(\text{BER})$}, 
legend pos=south east, 
legend cell align=left,
legend columns=5, 
xmin = 0,
xmax = 12,
ymin = -6,
ymax = 0,
ytick={0, -1, -2, -3, -4, -5, -6},
legend style={
at={(1.0, 1.27)},
anchor=north east,font=\tiny}
]

\addplot[line width=1pt][color=gray, solid, every mark/.append style={solid},mark=star,mark repeat = 2] table[x index =0, y index =1] {SimData/FP_ESI_Ntx4_Nc1024_Nsym256_R12_CI1_CK0_CD2_Nbps2_Nprea4_Npi16.dat};
\addlegendentry{{\footnotesize ESI}}

\addplot[line width=1pt][color=gray, solid, every mark/.append style={solid},mark=square,mark repeat = 3] table[x index =0, y index =1] {SimData/FP_DSI_Ntx4_Nc1024_Nsym256_R12_CI1_CK0_CD2_Nbps2_Nprea4_Npi16.dat};
\addlegendentry{{\footnotesize NeqDySI}}

\addplot[line width=1pt][color=black, solid] table[x index =0, y index =1] {SimData/FP_SISO_Ntx4_Nc1024_Nsym256_R12_CI1_CK0_CD2_Nbps2_Nprea4_Npi16.dat};
\addlegendentry{{\footnotesize SISO}}

\addplot[line width=1pt][color=lightgray, solid, every mark/.append style={solid},mark=+,mark repeat = 3] table[x index =0, y index =1] {SimData/FP_RDMult_Ntx4_Nc1024_Nsym256_R12_CI1_CK0_CD2_Nbps2_Nprea4_Npi16.dat};
\addlegendentry{{\footnotesize RDMult}}

\addplot[line width=1pt][color=black, solid, every mark/.append style={solid},mark=o,mark repeat = 2] table[x index =0, y index =1] {SimData/FP2_DDM_Ntx4_Nc1024_Nsym256_R12_CI1_CK0_CD2_Nbps2_Nprea4_Npi16.dat};
\addlegendentry{{\footnotesize DDM}}

\addplot[line width=1pt][color=gray, dashed, every mark/.append style={solid},mark=star,mark repeat = 2] table[x index =0, y index =1] {SimData/FP_ESI_Ntx4_Nc1024_Nsym256_R12_CI1_CK0_CD1_Nbps2_Nprea4_Npi16.dat};

\addplot[line width=1pt][color=gray, dashed, every mark/.append style={solid},mark=square,mark repeat = 3] table[x index =0, y index =1] {SimData/FP_DSI_Ntx4_Nc1024_Nsym256_R12_CI1_CK0_CD1_Nbps2_Nprea4_Npi16.dat};

\addplot[line width=1pt][color=black, dashed] table[x index =0, y index =1] {SimData/FP_SISO_Ntx4_Nc1024_Nsym256_R12_CI1_CK0_CD1_Nbps2_Nprea4_Npi16.dat};

\addplot[line width=1pt][color=lightgray, dashed, every mark/.append style={solid},mark=+,mark repeat = 3] table[x index =0, y index =1] {SimData/FP2_RDMult_Ntx4_Nc1024_Nsym256_R12_CI1_CK0_CD1_Nbps2_Nprea4_Npi64.dat};

\addplot[line width=1pt][color=black, dashed, every mark/.append style={solid},mark=o,mark repeat = 2] table[x index =0, y index =1] {SimData/FP_DDM_Ntx4_Nc1024_Nsym256_R12_CI1_CK0_CD1_Nbps2_Nprea4_Npi64.dat};

\end{axis}
\end{tikzpicture}
\caption{BER curves for the case of perfect channel knowledge, code rate $r=1/2$, and for perfect synchronization (solid) and for imperfect synchronization (dashed).
\label{fig:BER_synchronization} }
\end{center}
\end{figure}

Now, the channels are perfectly known, but the synchronization is performed using pilot subcarriers rather than having perfect synchronization. The \ac{siso} \ac{ofdm} system and the \ac{mimo} \ac{ofdm} systems utilizing \ac{esi} and \ac{dsi} employ $N_\text{p}=16$ pilot subcarriers. The \ac{mimo} \ac{ofdm} systems with \ac{ddm} and \ac{rdm} employ $N_\text{p}=64$ pilot subcarriers in order to ensure the same effective \ac{snr} as argued previously. The \ac{ber} curves visualized in Fig.~\ref{fig:BER_synchronization} show that the loss in \ac{ber} performance is minor for all considered systems.

\subsubsection{Perfect Channel Knowledge; Perfect CPE Synchronization; Disabled ICI}

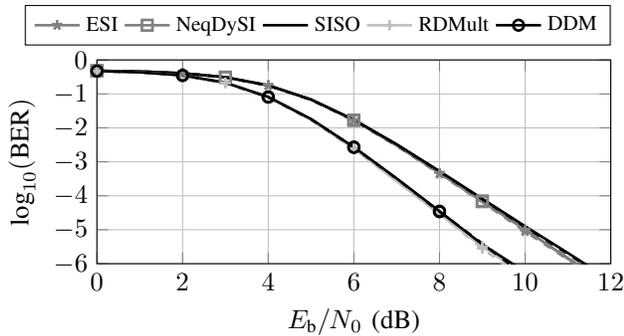
\begin{figure}[!t]
\begin{center}
\begin{tikzpicture}
\begin{axis}[compat=newest, 
width=.98\columnwidth, height = .5\columnwidth, grid, xlabel={$E_\text{b}/N_0$ (dB)}, 
ylabel={$\text{log}_{10}(\text{BER})$}, 
legend pos=south east, 
legend cell align=left,
legend columns=5, 
xmin = 0,
xmax = 12,
ymin = -6,
ymax = 0,
ytick={0, -1, -2, -3, -4, -5, -6},
legend style={
at={(1.0, 1.27)},
anchor=north east,font=\tiny}
]

\addplot[line width=1pt][color=gray, solid, every mark/.append style={solid},mark=star,mark repeat = 2] table[x index =0, y index =1] {SimData/FP_ESI_Ntx4_Nc1024_Nsym256_R12_CI1_CK0_CD2_Nbps2_Nprea4_Npi16.dat};
\addlegendentry{{\footnotesize ESI}}

\addplot[line width=1pt][color=gray, solid, every mark/.append style={solid},mark=square,mark repeat = 3] table[x index =0, y index =1] {SimData/FP_DSI_Ntx4_Nc1024_Nsym256_R12_CI1_CK0_CD2_Nbps2_Nprea4_Npi16.dat};
\addlegendentry{{\footnotesize NeqDySI}}

\addplot[line width=1pt][color=black, solid] table[x index =0, y index =1] {SimData/FP_SISO_Ntx4_Nc1024_Nsym256_R12_CI1_CK0_CD2_Nbps2_Nprea4_Npi16.dat};
\addlegendentry{{\footnotesize SISO}}

\addplot[line width=1pt][color=lightgray, solid, every mark/.append style={solid},mark=+,mark repeat = 3] table[x index =0, y index =1] {SimData/FP_RDMult_Ntx4_Nc1024_Nsym256_R12_CI1_CK0_CD2_Nbps2_Nprea4_Npi16.dat};
\addlegendentry{{\footnotesize RDMult}}

\addplot[line width=1pt][color=black, solid, every mark/.append style={solid},mark=o,mark repeat = 2] table[x index =0, y index =1] {SimData/FP2_DDM_Ntx4_Nc1024_Nsym256_R12_CI1_CK0_CD2_Nbps2_Nprea4_Npi16.dat};
\addlegendentry{{\footnotesize DDM}}

\addplot[line width=1pt][color=gray, dashed, every mark/.append style={solid},mark=star,mark repeat = 2] table[x index =0, y index =1] {SimData/FP_ESI_Ntx4_Nc1024_Nsym256_R12_CI0_CK0_CD2_Nbps2_Nprea4_Npi16.dat};

\addplot[line width=1pt][color=gray, dashed, every mark/.append style={solid},mark=square,mark repeat = 3] table[x index =0, y index =1] {SimData/FP_DSI_Ntx4_Nc1024_Nsym256_R12_CI0_CK0_CD2_Nbps2_Nprea4_Npi16.dat};

\addplot[line width=1pt][color=black, dashed] table[x index =0, y index =1] {SimData/FP_SISO_Ntx4_Nc1024_Nsym256_R12_CI0_CK0_CD2_Nbps2_Nprea4_Npi16.dat};

\addplot[line width=1pt][color=lightgray, dashed, every mark/.append style={solid},mark=+,mark repeat = 3] table[x index =0, y index =1] {SimData/FP_RDMult_Ntx4_Nc1024_Nsym256_R12_CI0_CK0_CD2_Nbps2_Nprea4_Npi16.dat};

\addplot[line width=1pt][color=black, dashed, every mark/.append style={solid},mark=o,mark repeat = 2] table[x index =0, y index =1] {SimData/FP_DDM_Ntx4_Nc1024_Nsym256_R12_CI0_CK0_CD2_Nbps2_Nprea16_Npi16.dat};

\end{axis}
\end{tikzpicture}
\caption{BER curves for code rate $r=1/2$, perfect synchronization and perfect channel knowledge. Simulation results are shown for activated (solid) and deactivated (dashed) ICI. The curves for DDM and RDMult as well as the curves for SISO, ESI and NeqDySI lie almost on top of each other.   
\label{fig:BER_woICI} }
\end{center}
\end{figure}

The loss in \ac{ber} performance due to \ac{ici}-induced distortions is analyzed by disabling \ac{ici}, while the channel is assumed to be known and the \ac{cpe} is compensated perfectly. The resulting \ac{ber} curves in Fig.~\ref{fig:BER_woICI} indicate that the loss in \ac{ber} performance is negligible for all considered multiplexing methods with the chosen parametrization. 

\section{Conclusion}
\label{sec:Conclusio}
In this work, a novel \ac{mimo} \ac{ofdm} joint radar and communication system designed for \ac{ddm} was presented. This multiplexing method generates transmit signals that are separable along the velocity axis in the \ac{rvm}. A thorough investigation of the properties of \ac{ddm} for the radar sensing task and the communication task has been carried out in this work. For the radar sensing task, it turned out that \ac{ddm} features the same \ac{snr} performance as \ac{rdm} and \ac{esi}. Differences between these multiplexing methods include \begin{itemize}
\item the average power per active subcarrier,
\item the processing gain, and
\item the maximum unambiguous range and velocity.
\end{itemize}

The transmit signals generated by a \ac{mimo} \ac{ofdm} system using \ac{ddm} were analyzed for the communication task as well. We showed that a \ac{siso} channel sufficiently models the communication channel and that this \ac{siso} channel is heavily time-varying, entailing the necessity of counter measures. We proposed a communication system specifically designed to cope with the time-varying nature of the channel. This communication system includes methods for data estimation, synchronization, and channel estimation, whose performances were evaluated by means of \ac{ber} simulations.



%


%
%
%





\bibliographystyle{IEEEtran}

\bibliography{References_DDM}

%








\end{document}